\documentclass[pdflatex,sn-mathphys-ay]{sn-jnl}

\usepackage{graphicx}%
\usepackage{multirow}%
\usepackage{amsmath,amssymb,amsfonts}%
\usepackage{amsthm}%
\usepackage{float}
\usepackage{svg}
\usepackage{mathrsfs}%
\usepackage[title]{appendix}%
\usepackage{xcolor}%
\usepackage{textcomp}%
\usepackage{manyfoot}%
\usepackage{booktabs}%
\usepackage{algorithm}%
\usepackage{algorithmicx}%
\usepackage{algpseudocode}%
\usepackage{listings}%

\theoremstyle{thmstyleone}%
\theoremstyle{thmstyletwo}%

\theoremstyle{thmstylethree}%

\begin{document}

\title[Article Title]{Data-driven turbulence modelling for magnetohydrodynamic flows in annular pipes}


\author[1]{\fnm{Alejandro} \sur{Montoya Santamaría}}\email{alejandromonto@tudelft.nl}

\author*[1]{\fnm{Tyler} \sur{Buchanan}}\email{t.s.b.buchanan@tudelft.nl}
\equalcont{These authors contributed equally to this work.}

\author[2]{\fnm{Francesco} \sur{Fico}}\email{f.fico@lboro.ac.uk}
\equalcont{These authors contributed equally to this work.}

\author[1]{\fnm{Ivan} \sur{Langella}}\email{i.langella@tudelft.nl}
\equalcont{These authors contributed equally to this work.}

\author[1]{\fnm{Richard} \sur{P. Dwight}}\email{r.p.dwight@tudelft.nl}
\equalcont{These authors contributed equally to this work.}

\author[1]{\fnm{Nguyen Anh Khoa} \sur{Doan}}\email{n.a.k.doan@tudelft.nl}
\equalcont{These authors contributed equally to this work.}

\affil*[1]{\orgdiv{Faculty of Aerospace Engineering}, \orgname{Delft University of Technology}, \orgaddress{\street{Kluyverweg 1}, \city{Delft}, \postcode{2600GB}, \state{Zuid Holland}, \country{The Netherlands}}}

\affil[2]{\orgdiv{Aeronautical and Automotive Engineering}, \orgname{Loughborough University}, \orgaddress{ \city{Loughborough}, \postcode{LE11 3TU},  \country{United Kingdom}}}



\abstract{We present a data-driven approach to Reynolds-averaged Navier-Stokes turbulence closure modelling in magnetohydrodynamic (MHD) flows. In these flows the magnetic field interacting with the conductive fluid induces unconventional turbulence states such as quasi two-dimensional (2D) turbulence, and turbulence suppression, which are poorly represented by standard Boussinesq models.  Our data-driven approach uses time-averaged Large Eddy Simulation (LES) data of annular pipe flows, at different Hartmann numbers, to derive corrections for the $k$-$\omega$ SST model.  Correction fields are obtained by injecting time averaged LES fields into the MHD RANS equations, and examining the remaining residuals.  The correction to the Reynolds-stress anisotropy is approximated with a modified Tensor Basis Neural Network (TBNN). We extend the generalised eddy hypothesis with a traceless antisymmetric tensor representation of the Lorentz force to obtain MHD flow features, thus keeping Galilean and frame invariance while including MHD effects in the turbulence model. The resulting data-driven models are shown to reduce errors in the mean flow, and to generalise to annular flow cases with different Hartmann numbers from those of the training cases.}
\keywords{Magnetohydrodynamics, Data-driven turbulence modelling, RANS, Tensor Basis Neural Network, SHapley Additive exPlanations}

\maketitle

\section{Introduction}


A critical component in fusion reactors is the Liquid Metal Breeding Blanket (LMBB), which typically employs lithium-based fluids. This system circulates liquid metal to act as a coolant and facilitate tritium production (Rubel \citeyear{articlerubentritium}; Smolentsev  \citeyear{smolentsev2021physical}). The magnetic field used to contain the fusion plasma interacts with the liquid metal flow, generating electric currents. These currents, combined with the magnetic field, create Lorentz forces that act on the liquid metal flow (Mistrangelo et al. \citeyear{MISTRANGELO2021112795}). The movement of liquid metals under these conditions is governed by magnetohydrodynamics (MHD), which integrates Maxwell's equations into the Navier-Stokes equations. The presence of a magnetic field can significantly alter the flow behavior, inducing large pressure drops, unconventional turbulence states, turbulence suppression, and increased anisotropy (Fico et al. \citeyear{Fico2023}). Currently, this turbulence behaviour can be captured using Large Eddy Simulation (LES) and Direct Numerical Simulation (DNS) frameworks. Nonetheless, the high computational cost of these simulations makes these techniques impractical for industrial applications compared to Reynolds-averaged Navier-Stokes (RANS) simulations. 

There exist several modifications of traditional Linear Eddy Viscosity Models (LEVMs) to include MHD related effects such as turbulence suppression. For instance, MHD turbulence models by Kenjeres and Hanjalic (\citeyear{Kenjere1997}) and Zhang et al. (\citeyear{Zhang2019}) proposed adding magnetic source terms to the $k$ and $\epsilon$ transport equations of the low-Reynolds $k$-$\epsilon$ turbulence model. These terms are meant to represent the decay in $k$ due to magnetic damping of turbulence. The first to propose this solution were Ji et al. (\citeyear{Ji1997}), who used the magnetic interaction parameter $N$ as the characteristic damping parameter. Other models are based on the same formulation, but use different model coefficients and a different characteristic damping parameter instead of $N$ (Kenjeres and Hanjalic \citeyear{Kenjere1997}; Zhang et al. \citeyear{Zhang2019}). However, while for MHD flows a tuned linear eddy viscosity model could potentially give a more accurate approximation of the turbulence kinetic energy, it cannot capture the turbulence anisotropy characteristics (Pope \citeyear{Pope1975}).

In parallel to the aforementioned development of turbulence closures for MHD flows, data-driven techniques using Machine Learning (ML) have shown promising results in deriving new turbulence closures through the use of higher fidelity simulations as training data (Jiang et al. \citeyear{Jiang2021}; Kaandorp and Dwight \citeyear{Kaandorp2020}; Ling et al. \citeyear{Ling2016}; Parish and Duraisamy \citeyear{Parish2016}; Cinnella \citeyear{cinnella2024data}).  This work relies on the generalised eddy viscosity hypothesis, presented by Pope (\citeyear{Pope1975}), in which the RST is modeled as a power-series in the mean strain-rate tensor and rotation-rate tensors. This power series has been used to represent the normalized turbulence anisotropy tensor $\boldsymbol{b}$ directly, or a non-linear correction ($\boldsymbol{b}^\Delta$) to a Boussinesq approximation of $\boldsymbol{b}$. In both cases the anisotropy is written as a tensor-valued function of resolved mean flow quantities, i.e. $\boldsymbol{b} \simeq f(\nabla \boldsymbol{U}, \boldsymbol{q})$, where $\boldsymbol{q}$ are flow features additional to the mean-velocity gradients. ML turbulence models have been trained for several engineering applications such as wind turbine wakes (Jigjid et al. \citeyear{jigjid2024simple}; Steiner et al. \citeyear{Steiner2020}) flows around 3D bluff bodies (Huijing et al. \citeyear{huijing2021data}), among others (Kaandorp and Dwight \citeyear{Kaandorp2020}; Wang et al. \citeyear{Wang2017}; Saez de Ocariz et al. \citeyear{Borde2021}). Given the promising results in these use cases, this study seeks to develop and evaluate an ML-based approach to derive a generalisable RANS turbulence model for MHD pipe flows.

We propose adding two correction terms to the $k$-$\omega$ SST turbulence model to account for MHD effects through the $k$-corrective frozen RANS approach of Schmelzer et al. (\citeyear{Schmelzer2020}). This method leverages high fidelity LES data to extract a correction term for the turbulence anisotropy, $\boldsymbol{b}^{\Delta}$, and a separate term for the deficit in the $k$-equation, $R$. To include MHD effects, we extend the generalised eddy hypothesis with a traceless antisymmetric tensor representation of the Lorentz force. The Tensor Basis Neural Network (TBNN) architecture by Ling et al. (\citeyear{Ling2016}) is then used to regress $\boldsymbol{b}^{\Delta}$. 
For $R$, we propose a modification of the TBNN architecture, the Scalar Basis Neural Network (SBNN) in which the basis tensors are substituted by scalar functions with dimensional units of $\mathrm{d}k/\mathrm{d}t$ $[\mathrm{m}^2\mathrm{s}^{-3}]$. The predicted corrections from these networks are then progressively substituted in the RANS equations using the methodology of Kaandorp and Dwight (\citeyear{Kaandorp2020}) to obtain an updated flow field in the annular pipe flow test cases.

The study is structured as follows. Section \ref{sec:dataset} presents the equations that govern the MHD RANS simulations as well as the modified $k$-$\omega$ SST turbulence model, including an explanation of the Frozen RANS method. The integration of the model in the RANS solver is discussed next, followed by an overview of the LES cases that have been used to train and validate the turbulence models.  Section \ref{sec:method} discusses the methodology for model regression, starting with the proposed extended eddy hypothesis for MHD flows  and followed by the regression techniques for the correction fields.  Section \ref{sec:results} proceeds to present the results of the \textit{a priori} testing of the model, followed by the \textit{a posteriori} testing. A summary of the main results and recommendation for future work are provided in the final section.











\section{MHD Simulation and Modelling Framework}
\label{sec:dataset}
We describe the equations of magnetohydrodynamics that are used in this study, the RANS turbulence modelling approach, and the high fidelity cases used as training data.  MHD flows can be characterized by two non-dimensional numbers, the Hartmann number $\mathrm{Ha}$, and the magnetic Reynolds number $\mathrm{Re}_m$:
\begin{equation}
\mathrm{Ha}=B_0 L \sqrt{\frac{\sigma}{\mu}}, \qquad \qquad
    \mathrm{Re}_m = \frac{U L}{\eta},
\end{equation}
where $U$ is the characteristic velocity of the flow, $L$ is the characteristic length, $\eta$ the magnetic diffusivity, $\sigma$ the electrical conductivity, $\mu$ the dynamic viscosity and $B_0$ is the magnetic field strength. The Hartmann number represents the ratio of the magnetic forces to viscous forces, while the magnetic Reynolds number is the ratio of the strength of the magnetic field induced by the movement of the conducting medium to the applied magnetic field (Lee and Choi \citeyear{Lee2001}).

\subsection{MHD RANS Equations}
\label{subsec:RANSeq}
The governing equations of MHD result from combining the Navier-Stokes and Maxwell equations. For liquid metal applications, the low magnetic Reynolds number assumption is applied, i.e.\ the magnetic field induced by electric current is negligible. Furthermore, the imposed magnetic field is assumed to be steady, and the fluid is considered incompressible with constant properties. Hence, by using Reynolds decomposition and applying Reynolds averaging, the MHD RANS equations for a liquid metal flow are obtained as
\begin{equation}
\label{eq:continuity}
    \nabla \cdot \boldsymbol{U}  = 0 ,
\end{equation}
\begin{equation}
\label{eq:momentum}
\rho\left[\frac{\partial \boldsymbol{U}}{\partial t}+(\boldsymbol{U} \cdot \nabla) \boldsymbol{U} \right]=-\nabla p +\rho \nu \nabla^2 \boldsymbol{U}+\boldsymbol{J} \times \boldsymbol{B_0} - \nabla \cdot \rho \overline{\boldsymbol{u'u'}} ,
\end{equation}
\begin{equation}
\label{eq:poisson}
    \Delta \varphi = \nabla \cdot ( \boldsymbol{U} \times \boldsymbol{B_0}) ,
\end{equation}
where $\boldsymbol{U}$ is the mean flow velocity vector and $ \boldsymbol{u'}$ is the vector of unsteady velocity fluctuations according to the Reynolds decomposition $\boldsymbol{u} = \boldsymbol{U} + \boldsymbol{u'}$. Hence, the Reynolds stress tensor is $\boldsymbol{\tau} := \overline{\boldsymbol{u'u'}}$, where the over-bar indicates ensemble averaging, requiring a turbulence closure. Furthermore, the Reynolds stress can be decomposed into its trace-free anisotropy $\boldsymbol{b}$ and the turbulence kinetic energy $k$ 
\begin{equation}
\label{eq:taukb}
    \boldsymbol{\tau} = 2k \left(\boldsymbol{b} + \frac{\boldsymbol{I}}{3} \right) .
\end{equation}
Here $p$ is the pressure, $\rho$ is the density, $\nu$ is the kinematic viscosity, $\boldsymbol{B_0}$ is the steady imposed magnetic field vector, $\varphi$ is the electrostatic potential and the current density $\boldsymbol{J}$ is given by 
\begin{equation}
\label{eq:ohmslaw2}
    \boldsymbol{J} = \sigma ( - \nabla \varphi + \boldsymbol{U} \times \boldsymbol{B_0}) ,
\end{equation}
\noindent
where the Lorentz force is defined as 
\begin{equation}
    \boldsymbol{F_L} = \frac{1}{\rho} \boldsymbol{J} \times \boldsymbol{B_0}.
\end{equation}

\subsection{MHD RANS Turbulence Closure Model}
\label{subsec:frozenRANS}

\begin{figure}[h]
 \centering
 \includegraphics[width=0.95\textwidth]{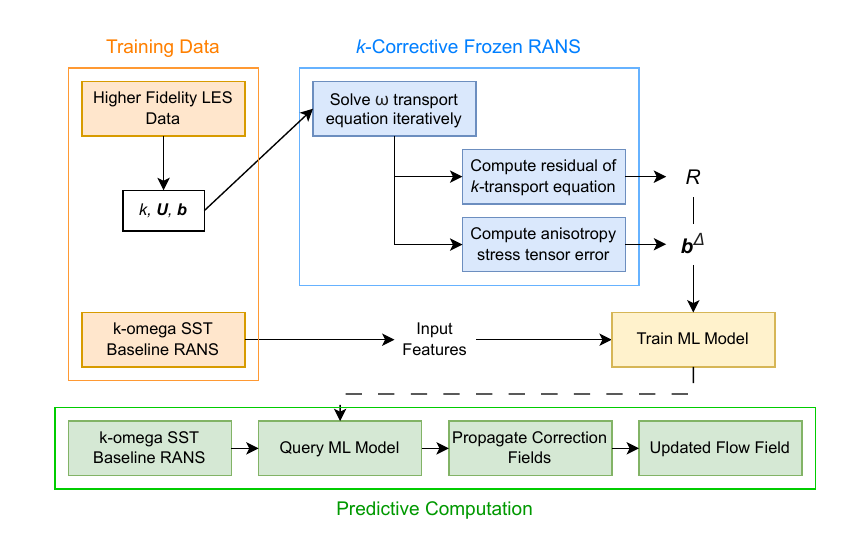}
 \caption{Proposed frozen RANS framework for data-driven turbulence modelling of MHD flows.}
 \label{fig:OL}
\end{figure}

 We use the $k$-$\omega$ SST eddy viscosity model (Menter \citeyear{menter1994two}) as the baseline for the MHD ML model. The $k$-$\omega$ SST model has two transport equations, one for $k$, and a second one for the specific turbulence dissipation rate $\omega$. It is based on the Boussinesq approximation of $\boldsymbol{b}$

\begin{equation}
\label{eq:bijboussinesq}
\hat{\boldsymbol{b}} = -\frac{\nu_t}{k} \boldsymbol{S},
\end{equation}
which assumes a linear relation between the mean flow strain rate $\boldsymbol{S} = (\nabla \boldsymbol{U} + \nabla \boldsymbol{U}^T) / 2$ and $\boldsymbol{b}$ through the scalar eddy viscosity $\nu_t$. To extract the MHD ML model from high fidelity data, we apply the $k$-corrective Frozen RANS method presented in Schmelzer et al. (\citeyear{Schmelzer2020}). This method adds two correction fields (i.e.\ functions of space) to the $k$-$\omega$ SST model: $\boldsymbol{b}^{\Delta}$ and $R$. The correction fields are extracted through frozen RANS simulations. In these simulations, LES mean-fields are injected into the RANS equations and kept frozen. Then, the $\omega$ transport equation is solved iteratively on the frozen velocity and kinetic energy fields. The first correction modifies the turbulence anisotropy
\begin{equation}
\label{eq:bijSSTML}
\hat{\boldsymbol{b}} = -\frac{\nu_t}{\hat{k}} \boldsymbol{S} + \boldsymbol{b}^{\Delta} .
\end{equation}
\noindent
The terms with the hat symbol $\hat{}$, such as $\hat{k}$ in (\ref{eq:bijSSTML}), are frozen to the LES values during the frozen RANS simulations. The second correction modifies the turbulence kinetic energy production. The $k$ and $\omega$ transport equations of the MHD ML model are 
\begin{align}
\frac{\partial \hat{k}}{\partial t}+\hat{\boldsymbol{U}} \cdot \nabla \hat{k} &= \hat{P_k} + R-\beta^* \hat{k} \omega+ \nabla \cdot \left[\left(\nu+\sigma_k \nu_t\right) \nabla \hat{k}\right], \label{eq:kSST} \\
\frac{\partial \omega}{\partial t}+\hat{\boldsymbol{U}} \cdot \nabla \omega&=\frac{\gamma}{\nu_t}\hat{P_k}-\beta \omega^2+\nabla \cdot \left[\left(\nu+\sigma_\omega \nu_t\right) \nabla \omega\right]+2\left(1-F_1\right) \sigma_{\omega 2} \frac{1}{\omega} \nabla \hat{k} \cdot \nabla \omega,\label{eq:omegaSST} .
\end{align}
Turbulence production $P_k$ in \eqref{eq:kSST} includes both Boussinesq and $\boldsymbol{b^{\Delta}}$ contributions to production:
\begin{equation}
\label{eq:pkfull}
    \hat{P_k} = \operatorname{min} \left( -2 k \left( \frac{\nu_t}{\hat{k}} \hat{\boldsymbol{S}} + \boldsymbol{b}^{\Delta} \right) : \nabla \hat{\boldsymbol{U}}, 10 \beta^* \omega k \right) .
\end{equation}
Finally, $\nu_t$ is defined as
\begin{equation}
    \nu_t = \frac{a_1 k}{\operatorname{max}(a_1 \omega, S F_2)} .
\end{equation}
With the frozen $\omega$ field resulting from the converged frozen RANS simulation, $R$ can be obtained from the $k$ transport equation (\ref{eq:kSST}), and $\boldsymbol{b}^{\Delta}$ from (\ref{eq:bijSSTML}). Furthermore, $a_1$, $\alpha$, $\beta$, $\beta^*$, $\gamma$, $\sigma_{\omega}$ and $\sigma_{\omega 2}$ are model constants, $F_1$ and $F_2$ are blending parameters and $S = \sqrt{2 \boldsymbol{S} : \boldsymbol{S}}$. A diagram summarising the $k$- corrective frozen RANS method is shown in Figure \ref{fig:OL}.

\subsection{Field Propagation}
In this study we choose to use an open-loop closure (Ling et al. \citeyear{Ling2016}), in which the ML model correction is evaluated only one time on the solution of the baseline SST model, and not updated thereafter.  This simplifies the integration and allows the use of more complex regression techniques with longer inference times.  This corresponds to the predictive computation phase in Figure \ref{fig:OL}. 

This approach requires completing a baseline RANS simulation (with $k$-$\omega$ SST) for each training case, using the same boundary-, and flow-conditions as the LES. The input features, the variables that are used to regress the correction fields, are computed using the data from these baseline RANS simulations. Similarly, when using the model in a simulation for a new CFD case, it is necessary to complete a baseline RANS simulation to acquire the input features needed to evaluate the ML model.

Once the model is queried, the corrections are inserted into the system of equations, and the simulation is restarted until the solution converged to a new velocity field. However, including the entire correction in a single iteration can cause numerical stability issues, depending on the magnitude of the corrections. Hence, the corrections are introduced gradually using a blending parameter $\gamma$, following the solution suggested by Kaandorp and Dwight (\citeyear{Kaandorp2020}),
\begin{equation}
     \gamma_i = \operatorname{max} \left( \operatorname{min} \left( \frac{i - i_\mathrm{start}}{i_\mathrm{end} - i_\mathrm{start}} , 1 \right) , 0 \right),
    \qquad \boldsymbol{b}^{\Delta}_i = \gamma_i \boldsymbol{b}^{\Delta},
    \qquad
    R_i = \gamma_i R,
\end{equation}
\noindent
where $i$ is the current iteration, $i_\mathrm{start}$ is the iteration at which blending begins and $i_\mathrm{end}$ is the iteration at which it ends.

\subsection{Flow case}
\begin{figure}[h]
 \centering
 \includegraphics[width=0.9\textwidth]{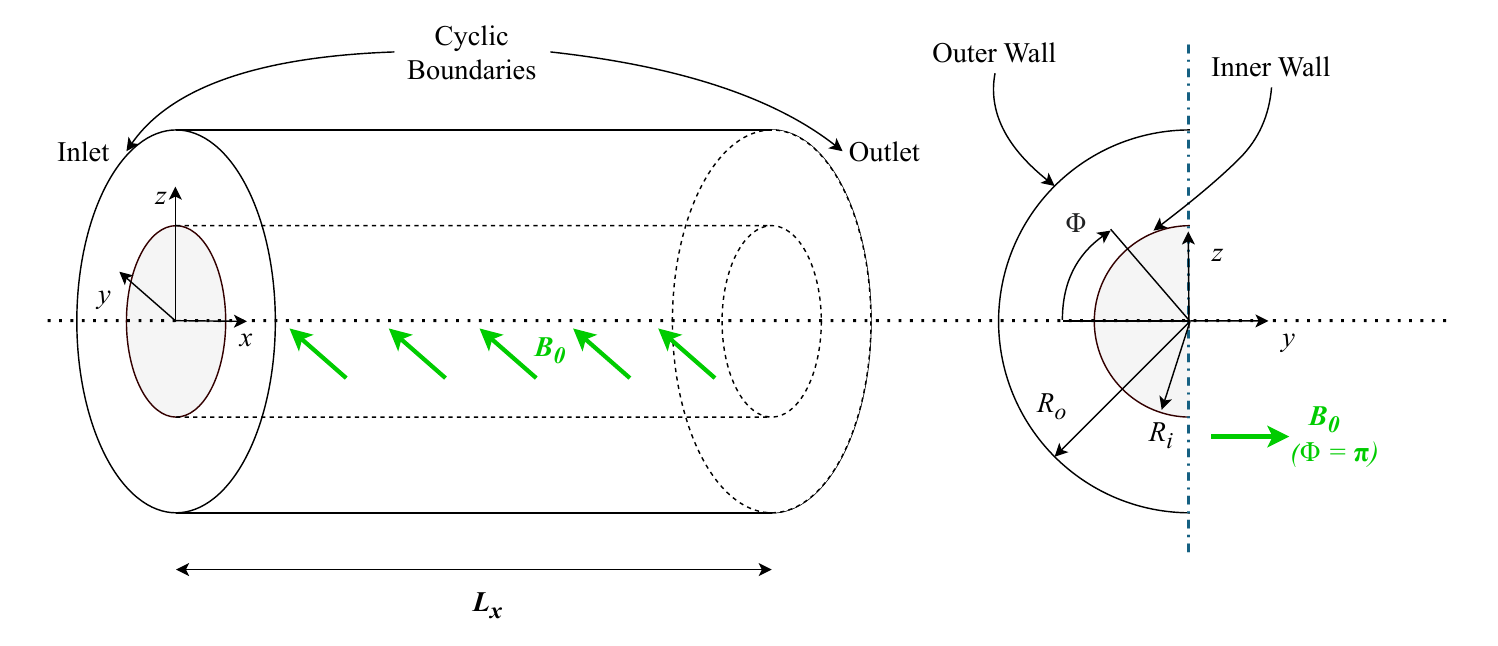}
 \caption{Cyclic duct domain  with a concentric annulus cross section and corresponding boundary conditions used for the LES. The bulk velocity is in the positive $x$ direction (Fico et al. \citeyear{Fico2023}).}
 \label{fig:BCLES}
\end{figure}

%
%
The high fidelity dataset consists of wall resolved LES simulations performed with OpenFOAM (Weller et al. \citeyear{openfoam}), employing a modified solver by Fico et al. (\citeyear{Fico2023}). The equations solved are the LES equivalent of \eqref{eq:continuity}, \eqref{eq:momentum} and \eqref{eq:poisson}. The study case is an annular pipe flow with cyclic boundary conditions at the inlet and the outlet of the numerical domain to simulate a fully developed boundary layer. 
The no-slip boundary condition is applied on the walls, which are also treated as electrically insulated. A constant uniform magnetic field of strength $B_0$ is applied in the transversal ($y$) direction.  Geometry and boundary conditions for the case are presented in Figure \ref{fig:BCLES}.  The mesh, geometry and flow parameters are summarised in Table \ref{tab:LESparams}, where the Reynolds number Re$_{D_h}$ is computed based on the bulk velocity of the pipe $U_b$ and the hydraulic diameter $D_h$, which for this geometry depends on the outer radius $R_o$ and the inner radius $R_i$. Furthermore, the length of the cyclic pipe $L_x$ is $6.25$ times $D_h$. The mesh is structured, and defined using the cylindrical coordinate system, where the radial coordinate is defined as $r = \sqrt{z^2 + y^2}$ and the azimuthal coordinate as $\phi = \arctan{\left(-z /y\right)}$. $N_x$, $N_r$ and $N_{\phi}$ are the number of cells in the longitudinal, radial and azimuthal directions respectively. $\Delta x^+$ and $\Delta r^+$ are the grid spacing in the longitudinal and radial directions. The minimum and maximum spacings in the radial direction are $\Delta r^+_{min}$ and $\Delta r^+_{max}$. The spacings in the azimuthal and longitudinal directions are uniform. Finally, the WALE model (Nicoud and Ducros \citeyear{nicoud1999subgrid}) is used to model the subgrid scales. A total of four cases have been simulated which differ only in their Hartmann number as indicated in Table~\ref{tab:LESparams}.

\begin{table}[h]
\centering
\caption{Physical, geometrical and mesh parameters used for the LES. The azimuthal width of the cells is constant.}
\label{tab:LESparams}
\begin{tabular}{@{}llll@{}}
\toprule
\multicolumn{2}{l}{Flow Conditions} & \multicolumn{2}{l}{Mesh Parameters}   \\ \midrule
$R_o$ & $2 R_i$         & $N_x$, $N_r$, $N_{\phi}$ & 260,80,520 \\ \midrule
$D_h$            & $\frac{4 \pi (R_o^2 -R_i^2)}{2 \pi (R_o + R_i)}$   & $\Delta r^+_{min}$       & 0.2        \\ \midrule
Re$_{D_h}$       & 8900           & $\Delta r^{+}_{max}$     & 5.3        \\ \midrule
Ha               & 0,40,60,120      & $\Delta x^+$             & 14         \\ \midrule
$L_x / D_h$      & 6.25             & \\ \bottomrule
\end{tabular}
\end{table}

\begin{figure}[h]
 \centering
 \includegraphics[width=\textwidth]{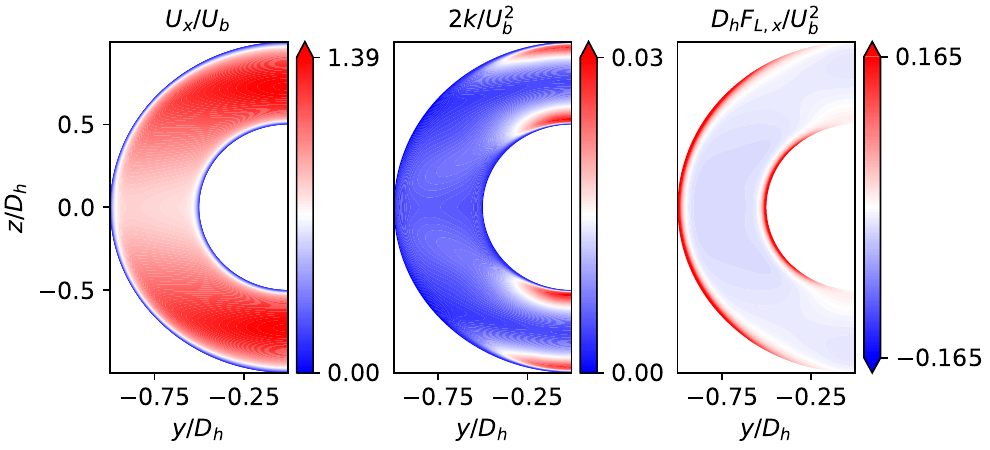}
 \caption{Non-dimensionalised mean flow longitudinal velocity, turbulence kinetic energy and longitudinal Lorentz force for LES of the $\mathrm{Ha}$= 60 case (Fico et al. \citeyear{Fico2023}).}
 \label{fig:LESfields}
\end{figure}

Figure~\ref{fig:LESfields} shows the normalised mean velocity field, turbulent kinetic energy and Lorentz force for a cross section of the co-annular duct.
Unlike an annular hydrodynamic pipe flow without magnetic forcing, the MHD mean flow is symmetric in the $z$-direction but not axisymmetric. In particular, the mean velocity in the plane $y = 0$ is significantly higher than in the plane $z = 0$, which is partly due to the stronger Lorentz force at locations near $z = 0$. For the same reason, the MHD effects on turbulence are stronger at $z=0$, resulting in turbulence suppression near the walls as one approaches $z=0$.

\section{Methodology: Data-driven model regression}
\label{sec:method}
We apply an extension of Pope's generalized eddy viscosity hypothesis to MHD flows using a methodology similar to that of Wu et al. (\citeyear{Wu2018}) to obtain the input features of the model. The target correction fields are approximated with two separate neural networks, following the TBNN architecture approach of Ling et al. (\citeyear{Ling2016}).


\subsection{Input Feature Selection}
\begin{figure}[h]
 \centering
 \label{fig:residuals}
 \includegraphics[width=0.9\textwidth]{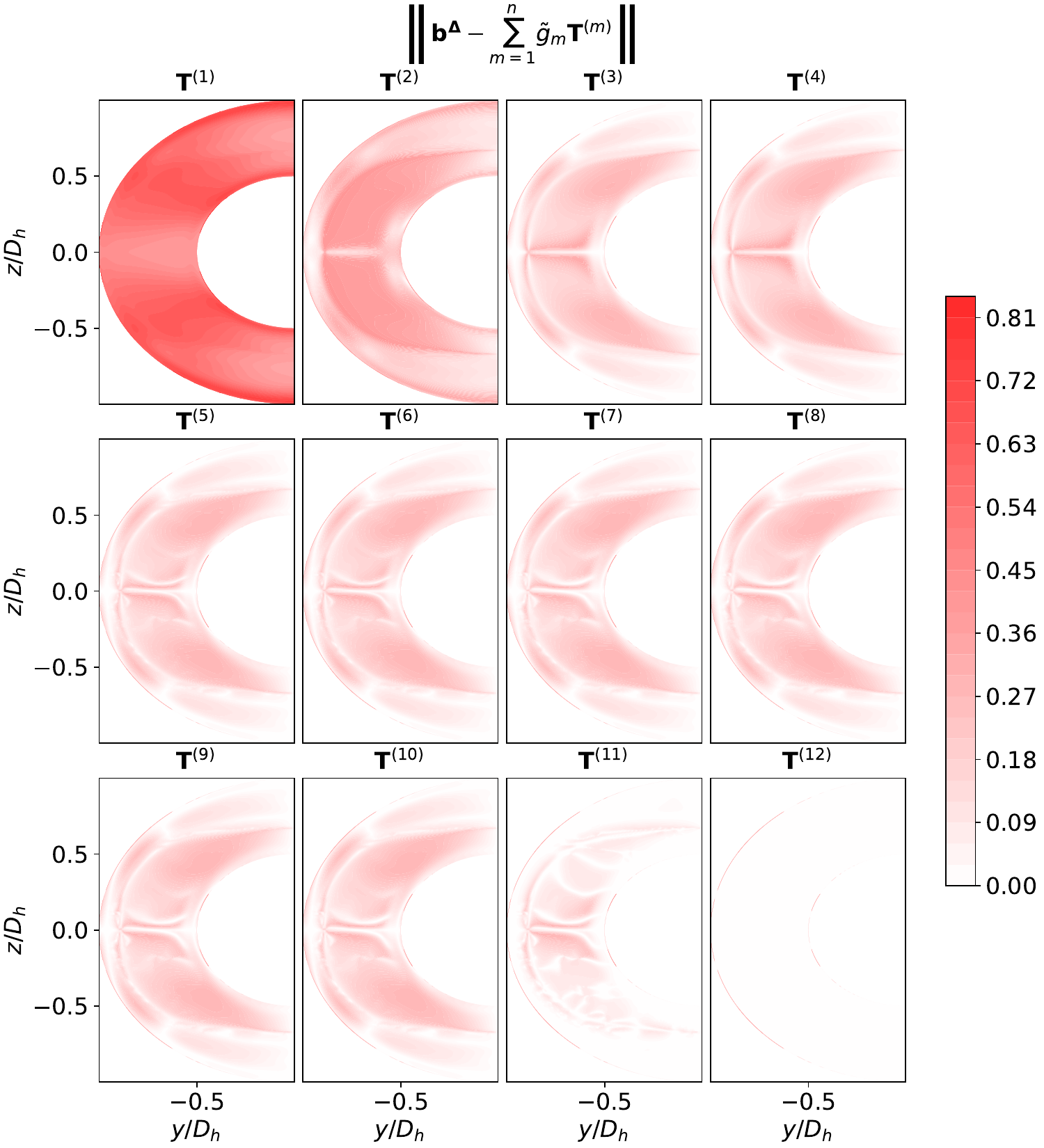}
 \caption{Magnitude of the residual of $\boldsymbol{b}^{\Delta}$ in the $\mathrm{Ha}$= 60 case after iteratively subtracting the projections of the candidate basis tensors.}
 \label{fig:residualprojections}
\end{figure}

Since the Navier-Stokes equations are Galilean and frame invariant, turbulence closures should have the same symmetries. One way of fulfilling these requirements is to use the generalised eddy viscosity hypothesis (Pope \citeyear{Pope1975}) to describe the correction. A key assumption is that the local mean-velocity gradients contain all the information relevant to the closure. The velocity gradient can be used to calculate the mean flow strain rate $\boldsymbol{S}$ and the mean flow rotation rate $\boldsymbol{\Omega} = (\nabla \boldsymbol{U} - \nabla \boldsymbol{U}^T) / 2$. Pope (\citeyear{Pope1975}) uses the turbulence timescale $t_\mathrm{turb} = k / \epsilon$ to make these tensors non-dimensional $\hat{\boldsymbol{S}} = t_\mathrm{turb} \boldsymbol{S}$, $\hat{\boldsymbol{\Omega}} = t_\mathrm{turb} \boldsymbol{\Omega}$, where $\epsilon$ is the turbulence dissipation rate. The anisotropy stress tensor is expressed as a function of these two non-dimensional tensors.  The most general form of a tensor-valued function of tensors is an (infinite) polynomial series, and the Cayley-Hamilton theorem limits the number of independent invariants and tensors, giving the finite expression
\begin{equation}
\label{eq:PopeEddy}
    \boldsymbol{b} = \sum_{n=1}^M g_n(I_1, I_2,...) \boldsymbol{T}^{(n)} .
\end{equation}
%
%
For 3 dimensional flows, there are a total of 10 basis tensors $\boldsymbol{T}^{(n)}$ and 5 tensor invariants $I_m$. This formulation is made for hydrodynamic flows.  To account for MHD effects on turbulence, it is necessary to at least include information on the Lorentz force $\boldsymbol{F_L}$ in the closure.  The approach suggested here is to include two additional antisymmetric tensors in the closed form expression  $\boldsymbol{A_k}$ and $\boldsymbol{A_L}$, such that the turbulence anisotropy can be written as
\begin{equation}
    \boldsymbol{b} = \boldsymbol{b} (\hat{\boldsymbol{S}} , \hat{\boldsymbol{\Omega}}, \hat{\boldsymbol{A_k}} , \hat{\boldsymbol{A_L}}),
\end{equation}
where, using the approach of Wang et al. (\citeyear{Wang2017}) we define 
\begin{equation}
    \boldsymbol{A_{L}} = - \boldsymbol{I} \times \boldsymbol{F_L},
    \qquad \qquad
    \boldsymbol{A_{k}} = - \boldsymbol{I} \times \boldsymbol{\nabla} k,
\end{equation}
such that $\boldsymbol{A_L}$ and $\boldsymbol{A_k}$ are traceless antisymmetric tensor representations of the Lorentz force and the kinetic energy gradient. This allows the generalised eddy hypothesis approach to account for the effects of these vectors without losing its Galilean and frame invariance. With these two additional tensors, a total of 47 linearly independent invariants can be derived using the same method as Wu et al. (\citeyear{Wu2018}). The complete list of these invariants can be found in Appendix \ref{secA2}. The addition of $\boldsymbol{A_{L}}$ and $\boldsymbol{A_{k}}$ also implies an increase in the number of basis tensors from the 10 derived by Pope.   To avoid having an excessive number of candidate basis tensors, only 5 additional tensors based on the Lorentz force are added:
\begin{equation}
\label{eq:tensorspope2}
\left.\begin{array}{ll}
\boldsymbol{T}^{(11)}=\hat{\boldsymbol{S}} \hat{\boldsymbol{A_L}}-\hat{\boldsymbol{A_L}} \hat{\boldsymbol{S}} & 
\boldsymbol{T}^{(12)}= \hat{\boldsymbol{A_L}} \hat{\boldsymbol{S}}^2 -\hat{\boldsymbol{S}}^2 \hat{\boldsymbol{A_L}} \\
\boldsymbol{T}^{(13)}= \hat{\boldsymbol{A_L}}^2-\frac{1}{3} \boldsymbol{I} \cdot \operatorname{Tr}\left(\hat{\boldsymbol{A_L}}^2\right) & \boldsymbol{T}^{(14)}= \hat{\boldsymbol{A_L}}^2 \hat{\boldsymbol{S}} + \hat{\boldsymbol{S}} \hat{\boldsymbol{A_L}}^2 \\
\boldsymbol{T}^{(15)}= \hat{\boldsymbol{A_L}} \hat{\boldsymbol{S}} \hat{\boldsymbol{A_L}}^2 - \hat{\boldsymbol{A_L}}^2 \hat{\boldsymbol{S}} \hat{\boldsymbol{A_L}} .
\end{array}\right\}
\end{equation}
We further reduce the number of basis tensors by removing those which are redundant. For this we apply the methodology proposed by Mandler and Weigand (\citeyear{Mandler2022}) to estimate coefficients for their generalized eddy viscosity model.  In their procedure the residual of the anisotropy correction is fit against each basis tensor sequentially (greedily) with least-squares.  The coefficients are given by the solution of the corresponding system of normal equations:
\begin{align}
\label{eq:mandler}
\tilde g_1 &= \frac{\boldsymbol{b}^{\Delta}: \boldsymbol{T}^{(1)}}{\boldsymbol{T}^{(1)}:\boldsymbol{T}^{(1)}} \\
\tilde g_n&=\frac{\left(\boldsymbol{b}^{\Delta} - \sum_{m=1}^{n-1} {\tilde g}_m \boldsymbol{T}^{(m)}\right): \boldsymbol{T}^{(n)}}{\boldsymbol{T}^{(n)}:\boldsymbol{T}^{(n)}}\qquad \text{for}\:n = 2,\dots, M,
\end{align}
where $M = 15$. This fit is performed pointwise for each cell in the computational mesh. Figure \ref{fig:residuals} presents the magnitude of the residuals of the Reynolds stress anisotropy after subtracting the projection of each tensor, which depends on its $\tilde g$ coefficient. Alternatively, these residuals can be described as the smallest possible error magnitude when using the basis tensors up to the one indicated in the plot.
%
%

Based on these results, it can be concluded that Pope's 10 basis tensors are not sufficient to describe the turbulence anisotropy tensor in these MHD pipe flow cases, as there remains a significant residual after the subtraction of the projections of these 10 tensors. The first three Pope tensors have a clear impact in reducing the residual, as well as the first two MHD based tensors. Therefore, the basis tensors that are used for the $\boldsymbol{b}^{\Delta}$ models are reduced to just $\boldsymbol{T}^{(1)}$, $\boldsymbol{T}^{(2)}$, $\boldsymbol{T}^{(3)}$, $\boldsymbol{T}^{(11)}$ and $\boldsymbol{T}^{(12)}$.

\subsubsection{Non-Dimensionalisation Factors}







We propose to use different non-dimensionalisation factors for computing the invariants and the basis tensors. For computing the invariants, the local turbulence time scale $t_\mathrm{turb}$ is used for $\boldsymbol{S}$ and $\boldsymbol{\Omega}$, as suggested by Pope (\citeyear{Pope1975}). For $\boldsymbol{A_L}$ and $\boldsymbol{A_k}$, the proposed non-dimensionalisation is
\begin{equation}
\label{eq:nondimLF}
    \hat{\boldsymbol{A_L}} = \frac{t_{turb} ^{3/2}}{\sqrt{\nu}} \boldsymbol{A_L},
    \qquad \qquad
    \hat{\boldsymbol{A_k}} = \frac{t_{turb}}{\sqrt{k}} \boldsymbol{A_k} .
\end{equation}
%

\begin{figure}[H]
 \centering
 \includegraphics[width=\textwidth]{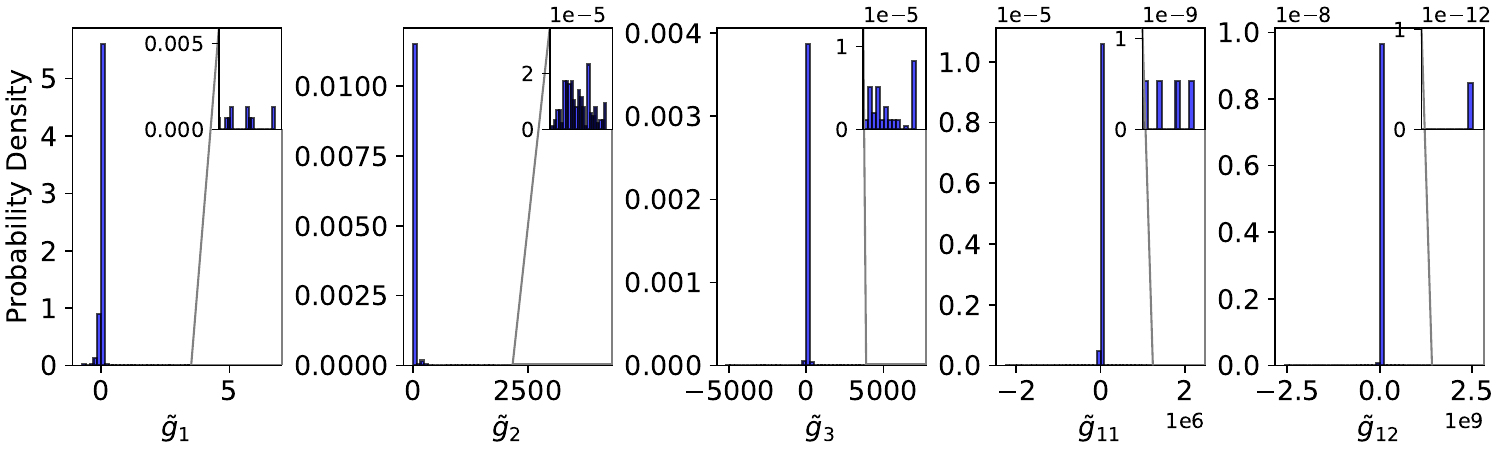}
 \caption{Distribution of the ${\tilde g}_n$ coefficients, with the basis tensors normalised by $t_{turb}$, for the $\mathrm{Ha}$ = 60 case. The inset axes in the top right of the plots zoom into section of the histogram corresponding to the highest coefficient values.}
 \label{fig:gnhstbad}
\end{figure}

\begin{figure}[H]
 \centering
 \includegraphics[width=\textwidth]{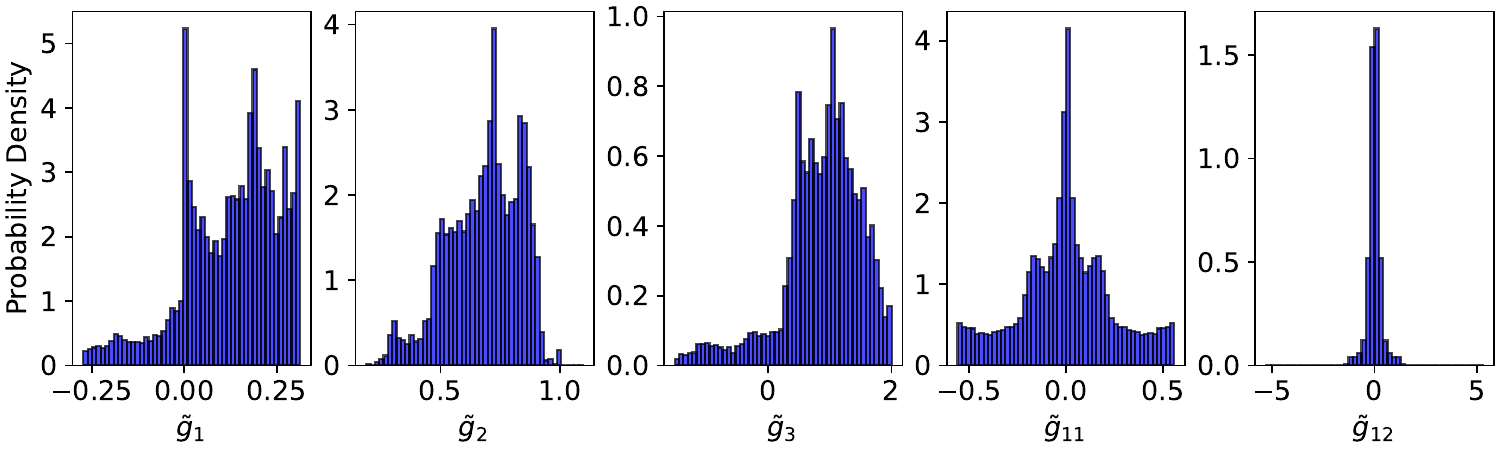}
 \caption{Distribution of the ${\tilde g}_n$ coefficients, with the basis tensors normalised by $t_{mean}$, for the $\mathrm{Ha}$ = 60 case.}
 \label{fig:gnhstgood}
\end{figure}

To compute the basis tensors, we instead use the mean flow timescale $t_{mean}$ as the non-dimensionalisation factor for $\boldsymbol{S}$ and $\boldsymbol{\Omega}$, a solution proposed by Miro et al. (\citeyear{Miro2023}). For $\boldsymbol{A_L}$, we use its $\ell^2$-norm
\begin{equation}
    \hat{\boldsymbol{S}} = \frac{1}{{\| \nabla \boldsymbol{U} \|}} \boldsymbol{S},
    \qquad \qquad
    \hat{\boldsymbol{A_L}} =  \frac{1}{\| \boldsymbol{A_L} \|} \boldsymbol{A_L} .
\end{equation}
\noindent
Using these non-dimensionalisation factors to compute the basis tensors, we obtain smooth distributions of the ${\tilde g}_n$ coefficients with values of order $1$. On the other hand, if the basis tensors are computed with the same non-dimensionalisation factors as the invariants, the value of the ${\tilde g}_n$ coefficients approaches $-\infty$ or $\infty$ when $t_\mathrm{turb}$ or the velocity gradient approach zero. For the MHD based tensors, the same issue appears when the magnitude of the Lorentz force approaches zero. This entails that the data distribution of the ${\tilde g}_n$ coefficients is more difficult to regress accurately. A comparison of the distributions is presented in Figures \ref{fig:gnhstbad} and \ref{fig:gnhstgood}.

\subsubsection{Additional Features}

In addition to 47 invariants, 12 additional features based on physical considerations are also considered. These are listed in Table \ref{tab:newfeatures}. The aim of adding these features is to alleviate the limitations that are imposed by the assumptions that the generalised eddy hypothesis makes. This has been done for data-driven closure models of non-MHD flows, of which Re$_t$ (Jiang et al. \citeyear{Jiang2021}), Re$_y$, $\nu_t / 100 \nu$ and $q_T$ (Schmelzer et al. \citeyear{Schmelzer2020}) are used here. The remaining are features of MHD, where $t_{turb} / t_{mag}$ comes from the work of Kenjeres and Hanjalic (\citeyear{Kenjere1997}) and the others make use of $\boldsymbol{A_L}$ or the Lorentz force gradient $\nabla \boldsymbol{F_L}$.

%

\begin{table}[h]
\centering
\caption{ Additional features added as inputs to the turbulence models which are not invariants of the tensor basis.}
\label{tab:newfeatures}
\begin{tabular}{@{}lll@{}}
\toprule
Description & Symbol & Equation \\ \midrule
\begin{tabular}[c]{@{}l@{}}Turbulence\\ Reynolds Number\end{tabular}            &  Re$_t$          & $\frac{k^2}{\nu \epsilon}$ \\ \midrule
\begin{tabular}[c]{@{}l@{}}Ratio of characteristic turbulence \\ timescale to magnetic braking time\end{tabular}                                            &                 & $\frac{t_{turb}}{t_{mag}}$   \\ \midrule
\begin{tabular}[c]{@{}l@{}}Scaled wall distance based \\ Reynolds number\end{tabular}                                                                                                                                                                                      & Re$_y$          & $2 - \operatorname{min}(\frac{\sqrt{k} y_{wall}}{50 \nu})   $  \\ \midrule
\begin{tabular}[c]{@{}l@{}}Scaled ratio of eddy viscosity to \\ laminar viscosity\end{tabular}                                                              &                 & $\frac{\nu_t}{100 \nu}$       \\ \midrule
\begin{tabular}[c]{@{}l@{}}Ratio of characteristic turbulence \\ timescale to mean flow timescale\end{tabular}                                              & $q_T$            & $\frac{t_{turb}}{t_{mean}}$    \\ \midrule
\begin{tabular}[c]{@{}l@{}}Ratio of mean flow timescale to \\ magnetic braking time\end{tabular}                                                            &                 & $\frac{t_{mean}}{t_{mag}}$      \\ \midrule
\begin{tabular}[c]{@{}l@{}}Ratio of the $l^2$ norms of the \\ Lorentz force tensor and the strain\\  rate tensor times $t_{turb}$\end{tabular}              & $q_{AS \omega}$ & $\frac{\| \frac{t_{turb}^{1.5}}{\sqrt{\nu}} \boldsymbol{A_L} \|}{\| t_{turb} \boldsymbol{S} \|}$      \\ \midrule
\begin{tabular}[c]{@{}l@{}}Ratio of the $l^2$ norms of the \\ Lorentz force tensor and the strain\\  rate tensor times $t_{mean}$\end{tabular}              & $q_{AS m}$      & $\frac{\| \frac{t_{turb}^{1.5}}{\sqrt{\nu}} \boldsymbol{A_L} \|}{\| t_{mean} \boldsymbol{S}\|}$        \\ \midrule
$l^2$ norm of the Lorentz force tensor                                                                                                                      & $q_{A}$         & $\| \frac{t_{turb}^{1.5}}{\sqrt{\nu}} \boldsymbol{A_L} \|$     \\ \midrule
\begin{tabular}[c]{@{}l@{}}Ratio of the square root $l^2$ norm of the\\  Lorentz force gradient and the $l^2$ norm\\  of the velocity gradient\end{tabular} & $q_{LS}$        & $\frac{\sqrt{\| \nabla \boldsymbol{F_L}}}{\nabla \boldsymbol{U}}$ \\ \midrule
\begin{tabular}[c]{@{}l@{}}Alignment of the Lorentz force and \\ velocity gradients\end{tabular}                                                            & $q_{\alpha LS}$ & $ \alpha( \nabla \boldsymbol{F_L}, \nabla \boldsymbol{U})$       \\ \midrule
\begin{tabular}[c]{@{}l@{}}Alignment of the Lorentz force and\\  velocity gradients without normalisation\end{tabular}                                      & $q_{g LS}$      & $t_{turb} \frac{\nabla \boldsymbol{F_L} : \nabla \boldsymbol{U}}{\| \nabla \boldsymbol{U} \|^2}$  \\ \bottomrule
\end{tabular}
\end{table}

%

\subsubsection{Scalar Basis for $R$}

A distinct scalar basis must be used for the $R$ correction field because it is a scalar quantity with dimensions of [m$^2$s$^{-3}$]. The method used to construct this scalar basis follows the approach applied in SpaRTA by Schmelzer et al. (\citeyear{Schmelzer2020}). The scalar basis can be derived from the tensor basis using Equation (\ref{eq:Gnformula}).
\begin{equation} 
\label{eq:Gnformula} 
G^{(n)} = 2 k \boldsymbol{T}^{(n)}: \boldsymbol{\nabla \boldsymbol{U}} ,
\end{equation}
\noindent
where
\begin{equation} 
\label{eq:Pkbasis} 
R = \sum_{n} c_n(I_1, I_2, \dots) G^{(n)} .
\end{equation}
\noindent
Analogously to $g_n$, the basis coefficients $c_n$ are a function of the invariants and the additional features.  Another scalar variable that can be included in the basis is the turbulence dissipation rate, $\epsilon$, which also has dimensions of [m$^2$s$^{-3}$].

Since the correction is scalar, even a single basis function could suffice for modelling $R$. However, allowing the model to utilize multiple basis functions enables the regression to approximate a collection of simpler functions rather than fitting one complex function. Additionally, in this context, the use of the turbulence time scale does not pose the same issues encountered with the tensor basis. As a result, tensors are computed using both the turbulence and mean timescales for non-dimensionalisation, producing a total of 21 candidate scalar basis functions. The functions that incorporate the turbulence timescale are labeled $G_t^{(n)}$. The base scalars derived from $\boldsymbol{A_L}$-based basis tensors were excluded because the resulting functions had magnitudes below 10$^{-10}$. Ultimately, only $G^{(1)}, G^{(6)}, G_t^{(1)}, G_t^{(6)}$ and $ \epsilon$ were selected for the regression, based on their significantly higher correlation with $R$ compared to the other candidate basis functions.

\subsection{Regression Techniques}

\begin{figure}[h]
 \centering
 \includegraphics[width=0.8\textwidth]{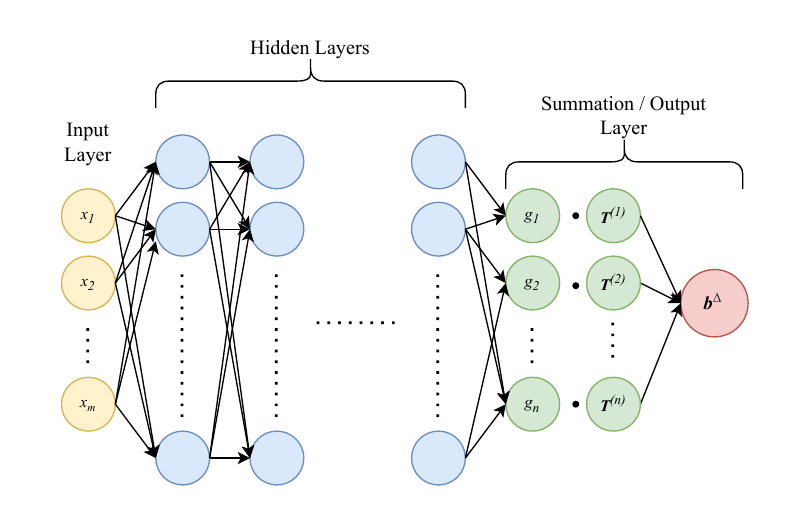}
 \caption{Schematic of the TBNN, where $m$ is the number of input features, and $n$ is the number of basis tensors or functions.}
 \label{fig:TBNNdiagram}
\end{figure}

Due to its capability to capture complex functions, the chosen regression technique for $\boldsymbol{b}^{\Delta}$ is the Tensor Basis Neural Network (TBNN), which was first proposed by Ling et al. (\citeyear{Ling2016}), and later implemented using the pytorch library (Paszke et al. \citeyear{paszke2019pytorch}) by Parashar et al. (\citeyear{parashar2020modeling}). The TBNN embeds the frame and Galilean invariance of the generalised eddy hypothesis using the architecture of the neural network. The TBNN consists of a Fully Connected Feed Forward network (FCFF), which outputs the basis coefficients $g_n$ using a selected set of input features, including basis invariants and additional features presented in Table \ref{tab:newfeatures}. The next layer multiplies the basis coefficients by the corresponding basis tensors $\boldsymbol{T}^{(n)}$, and the final layer performs a summation to output the prediction of $\boldsymbol{b^{\Delta}}$. This is presented visually in Figure \ref{fig:TBNNdiagram}.

To alleviate overfitting issues, we use dropout layers with probability $p = 0.5$ on all hidden layers of the FCFF, and $p = 0.2$ on the input layer, where $p$ is the probability of each neuron of a layer being removed in each forward pass. The activation function is the hyperbolic tangent, a bounded function, to avoid large errors when extrapolating. The size of the FCFF is based on the number of available data points $N$ from the CFD cases used for training, ensuring that the number of trainable parameters in the network does not exceed 10 $\%$ of $N$. This resulted in 7 hidden layers with 34 neurons per layers when using all the 4 cases for training, and 6 hidden layers with 31 neurons per layer when using 3 cases for training.


80 $\%$ of the data is used for training and the rest is used for validation. The training is stopped once the loss on the validation data does not decrease over the last 40 epochs of the training. In each epoch, data is processed in batches of 32, starting with a learning rate of 10$^{-4}$, which is halved after 10 epochs without a decrease in training loss. We use the Adam optimiser (Kingma and Ba \citeyear{kingma2014adam}) to train the model, with Mean Squared Error (MSE) as the loss function.

The proposed solution for regressing $R$ is an adaptation of the TBNN, the Scalar Basis Neural Network (SBNN). The architecture of this network is practically identical to the TBNN, with the single adaptation of changing the layer after the FCFF by multiplying the output coefficients of the FCFF by the scalar functions $G^{(n)}$ instead of the basis tensors. Furthermore, for the SBNN the activation function is changed to the Gaussian Error Linear Unit (GeLU), and no dropout layers are used.

\subsection{Test Matrix}

To maximize the information gained from both types of testing, rather than selecting a single test case, we train the model on three different combinations of the MHD cases, all of which include the $\mathrm{Ha}$ = 0 case. The rationale for this is that because in the pure hydrodynamic case, the Lorentz force-based input features are zero, by always including this case in the training, the model can learn to differentiate between phenomena typical of hydrodynamic flows and those arising from MHD effects. Furthermore, since this model is not intended for use in standard hydrodynamic flows, testing its extrapolation to $\mathrm{Ha}$ = 0 is unnecessary. The model trained on all four cases is referred to as the "All Data" model in the following chapters and is evaluated on the three MHD cases to estimate the best results possible using the given methodology. The models trained on three out of the four cases are called test models and are evaluated on their respective test cases. The final test matrix is shown in Table \ref{tab:testmatrix}.

\begin{table}[h]
\centering
\caption{ Test matrix for the \textit{a priori} and \textit{a posteriori} testing of the model.}
\label{tab:testmatrix}

\begin{tabular}{@{}llll@{}}
\toprule
Model Type    & $\mathrm{Ha}$ of Training Cases & $\mathrm{Ha}$ of Test Cases & $\mathrm{Ha}$ of Propagation Cases \\ \midrule \midrule
All Data      & 0,40,60,120          & N.A.             & 40,60,120               \\ \midrule
Ha = 40 Test  & 0,60,120             & 40               & 40                      \\ \midrule
Ha = 60 Test  & 0,40,120             & 60               & 60                      \\ \midrule
Ha = 120 Test & 0,40,60              & 120              & 120                     \\ \midrule \bottomrule
\end{tabular}
\end{table}

\section{Results and Discussion}
\label{sec:results}

The testing of the MHD ML turbulence model is divided into \textit{a priori} and \textit{a posteriori} testing. \textit{A priori} tests compare the output of the models to the target correction fields for the different cases. On the other hand, in \textit{a posteriori} testing the data-driven turbulence model is implemented into a RANS solver and the results of the simulated test cases are compared to those of the equivalent high fidelity LES case.

\subsection{\textit{A priori} Testing}
Firstly, the results for the $\boldsymbol{b}^{\Delta}$ model are discussed, followed by the results for the $R$ model. 

\subsubsection{Turbulence Anisotropy Correction}
\label{subsec:tbnnapriori}

An overview of the $\boldsymbol{b}^{\Delta}$ prediction for the four different Hartmann number cases is shown in Table \ref{tab:bdelta}. The results are presented for four models trained on different combinations of the available CFD cases. The error is computed using Root Mean Squared Error (RMSE), defined for $\boldsymbol{b}^{\Delta}$ as shown in Equation (\ref{eq:rmsebijj}), where $N$ is the number of data points included in the operation

\begin{equation}
\label{eq:rmsebijj}
    \operatorname{RMSE}(\boldsymbol{b}^{\Delta}) = \sqrt{\frac{1}{9N} \sum_{m = 1}^N \sum_{i = 1}^3 \sum_{j = 1}^{3} (b_{i j, out}^{\Delta(m)} - b_{i j, DNS}^{ \Delta (m)})^2} .
\end{equation}

\begin{table}[b]
\centering
\caption{RMSE of $\boldsymbol{b}^{\Delta}$ of the TBNN models for each CFD case. As reference, the RMSE of the target correction is given in the first row (Case $\boldsymbol{b}^{\Delta}=0$).}
\label{tab:bdelta}
\begin{tabular}{@{}lllll@{}}
\cmidrule(l){2-5}
                                           & \multicolumn{4}{l}{RMSE($\boldsymbol{b}^{\Delta}$)}                                                                                \\ \midrule\midrule
Case & $\mathrm{Ha}$ = 0 & $\mathrm{Ha}$ = 40 & $\mathrm{Ha}$ = 60 & $\mathrm{Ha}$ = 120 \\ \midrule\midrule
$\boldsymbol{b}^{\Delta}$ = 0 & 0.161 & 0.202  & 0.192  & 0.195   \\ \midrule
All Data Model & 0.0242 & 0.0497  & 0.0648  & 0.122    \\ \midrule
Ha = 40 Test Model    & 0.0297 & 0.0589  & 0.0648  & 0.117   \\ \midrule
Ha = 60 Test Model    & 0.0252 & 0.0500  & 0.0718  & 0.121   \\ \midrule
Ha = 120 Test Model     & 0.0165 & 0.0431  & 0.0668  & 0.155    \\ \midrule\bottomrule
\end{tabular}
\end{table}

\begin{figure}[H]
 \centering
 \includegraphics[width=\textwidth]{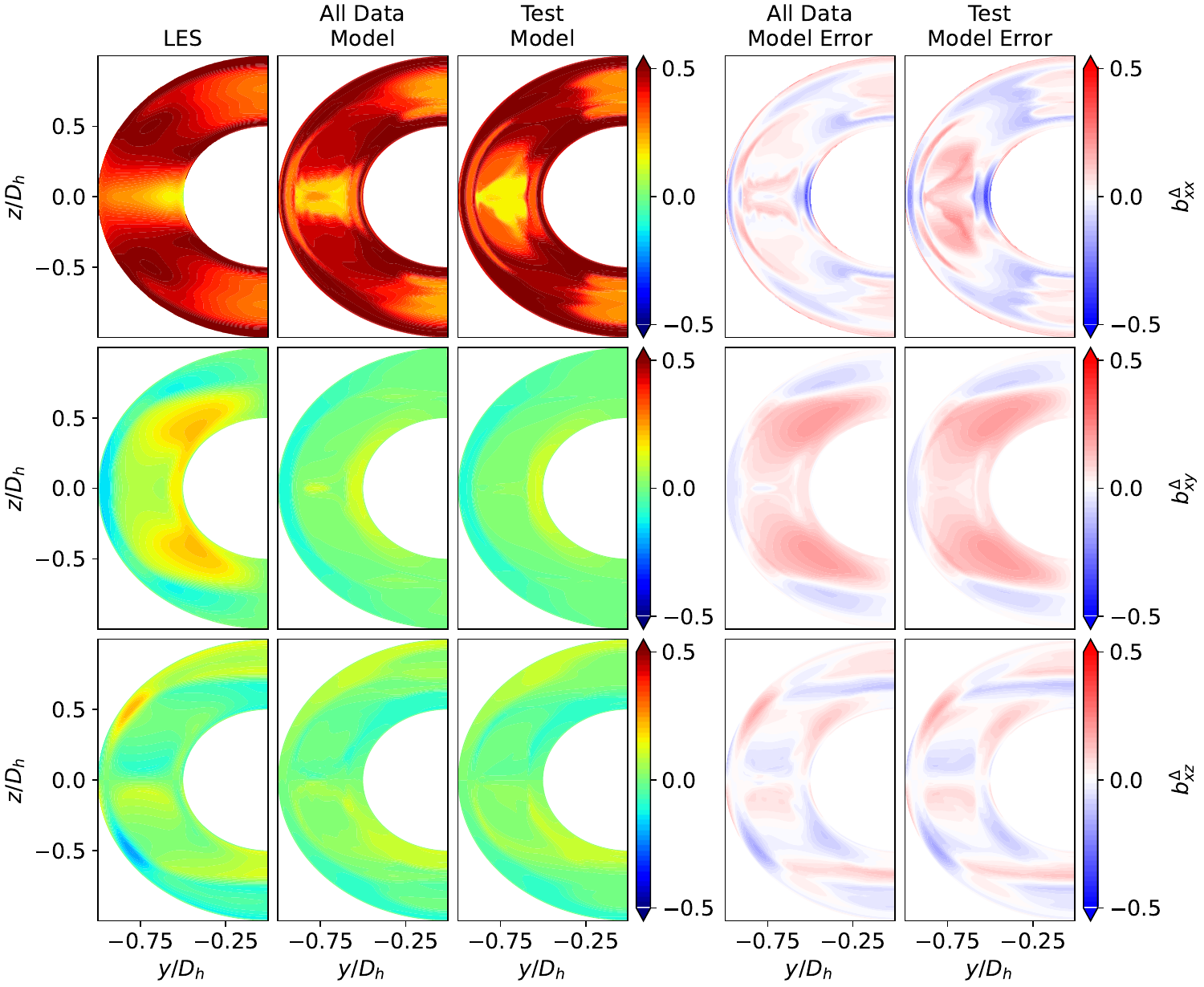}
 \caption{Transversal section contours of the $\boldsymbol{b}^{\Delta}$ prediction and error of the TBNN All Data and test models on the $\mathrm{Ha}$ = 60 case.}
 \label{fig:ha60_full}
\end{figure}

The results in Table \ref{tab:bdelta} show that the trained ML models achieve a considerable reduction in RMSE with respect to the eddy viscosity baseline ($\boldsymbol{b}^{\Delta}$ = 0). However, there exists a trend of increasing error with increasing Hartmann number. Relative improvements with respect to baseline are 71$\%$, 63$\%$ and 21$\%$ for the $\mathrm{Ha}$ = 40, 60, 120 test models respectively. This is expected partially because the differences between the mean velocity field from the RANS baseline simulation (eddy viscosity model)  and from the LES are also observed to become larger with the Hartman number. Moreover, the test models perform worse than the `All Data model' for their corresponding test case. This is especially notable for the $\mathrm{Ha}$ = 120 test model, where all its training cases correspond to lower $\mathrm{Ha}$ values. As a result, the flow features in the test case frequently fall outside the range encountered during training, requiring the model to extrapolate when making predictions. The error can be analysed in more depth by observing its distribution across the cross section of the pipe for each of the components of $\boldsymbol{b}$. Figure \ref{fig:ha60_full} shows the values and error of the turbulence anisotropy components present in the $x$-momentum equation for the $\mathrm{Ha}$ = 60 case. The general trends for all components are captured, except near the wall for $b_{xx}^{\Delta}$ at $\phi = 0$, and at $\phi = \pi / 4$ and $\phi = - \pi / 4$ for $b_{xy}^{\Delta}$. For both of these locations, the ML model fails to capture the behaviour of these Reynolds stress components. Similar results are also observed for the other available cases at $\mathrm{Ha}$ = 40 and $\mathrm{Ha}$ = 120, indicating that the ML model is missing features to accurately describe turbulence in these regions of the MHD flow.

\begin{figure}[b]
 \centering
 \includegraphics[width=\textwidth]{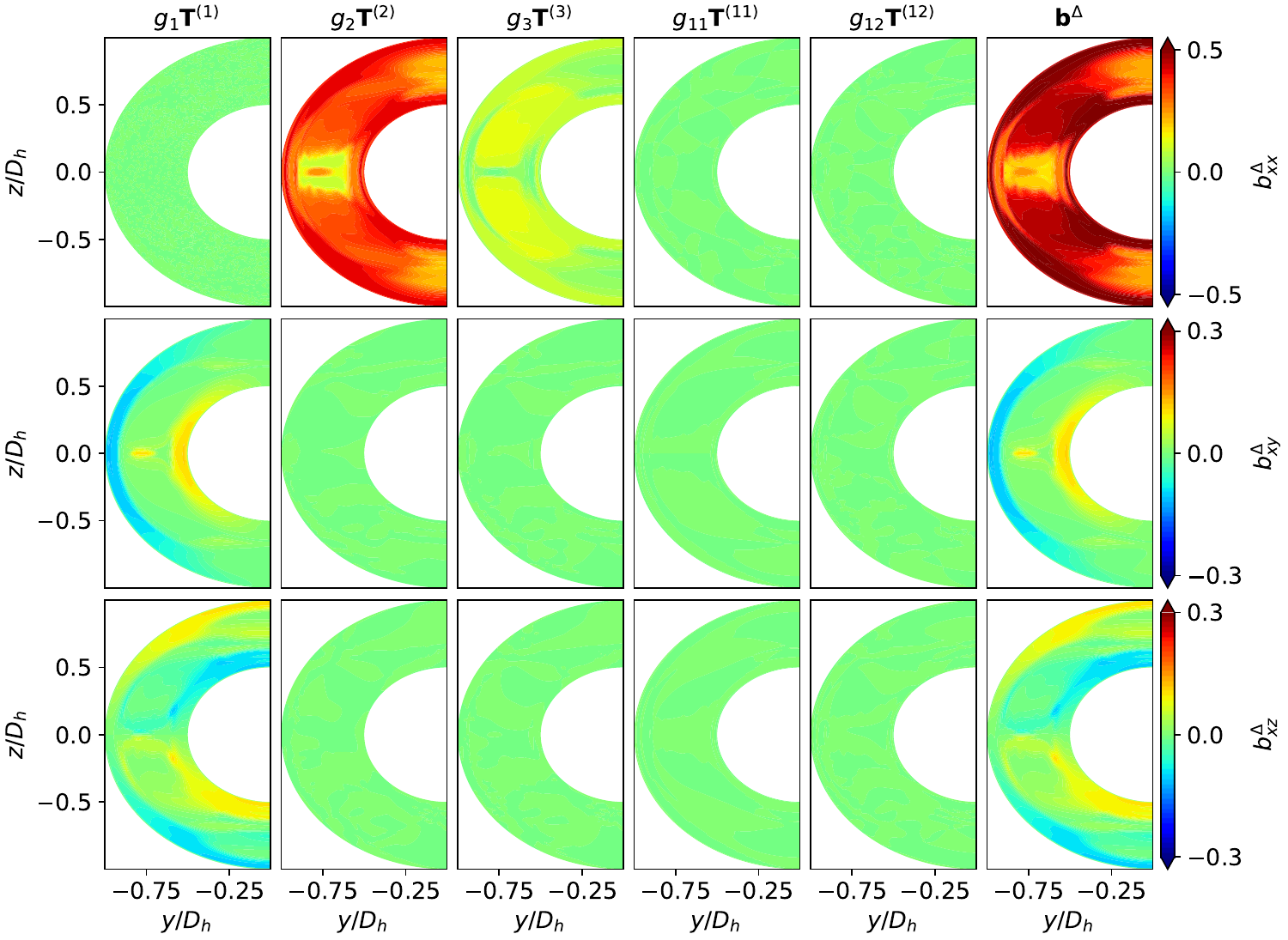}
 \caption{Transversal section contours of the $\boldsymbol{b}^{\Delta}$ contribution from each basis tensor using the All Data model for the $\mathrm{Ha}$ = 60 case.}
 \label{fig:contributions}
\end{figure}

The contributions of each pair of basis tensors and their respective coefficients can be computed individually, providing insight into which tensors have the most influence on the final output of $\boldsymbol{b}^{\Delta}$ and for which specific components. An example of this is illustrated for the $\mathrm{Ha}$ = 60 case using the All Data model in Figure \ref{fig:contributions}. One key takeaway from this figure is the clear inadequacy of the eddy viscosity assumption for these types of flows, as $\boldsymbol{T}^{(1)} = \hat{\boldsymbol{S}}$ fails to account for the contributions to the normal stress component. Another important observation is that the two MHD tensors, $\boldsymbol{T}^{(11)}$ and $\boldsymbol{T}^{(12)}$, have minimal influence on the model’s output across all six components. As shown in Figure \ref{fig:contributions}, while these tensors can help reduce the error in predicting $\boldsymbol{b}^{\Delta}$ given the ${\tilde g}_n$ coefficient values (see Equation (\ref{eq:mandler})), the available input features do not provide the model with sufficient information to accurately estimate these ${\tilde g}_n$ coefficients with the proposed TBNN regression. As a result, the model relies solely on the three lowest-order Pope tensors to make its predictions, making them crucial for producing a reasonable estimate of $\boldsymbol{b}^{\Delta}$ in MHD annular flows. The first tensor is the primary contributor to the shear components $xy$ and $xz$, while the second and third tensors add up to the normal component $xx$. 

\begin{figure}[tbh]
 \centering
 \includegraphics[width=0.7\textwidth]{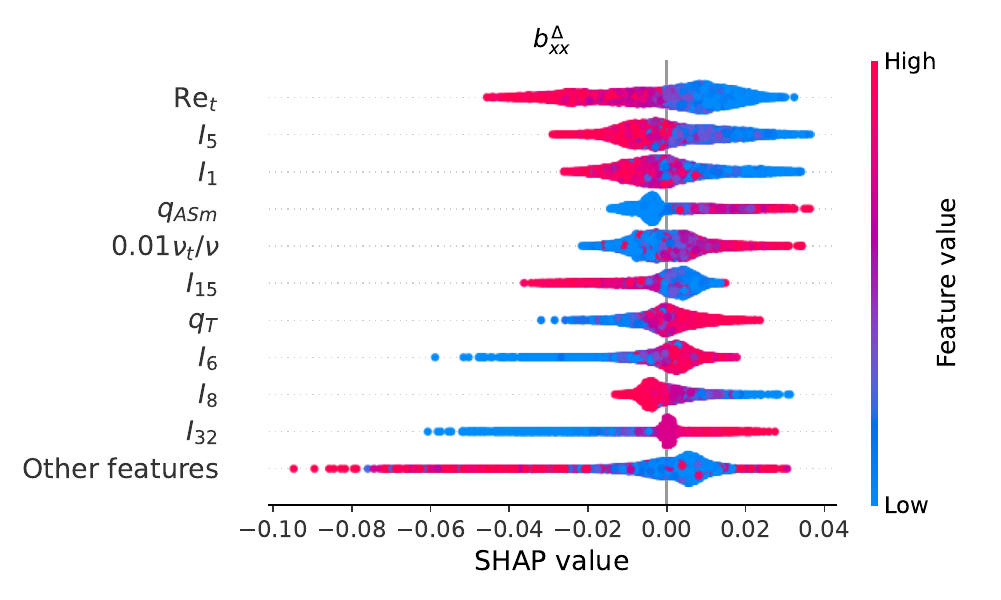}
 \caption{SHAP values for the input features of the SBNN with respect to $\boldsymbol{b}^{\Delta}$. The 10 features with the highest influence are shown separately in descending order.}
 \label{fig:shapbxx}
\end{figure}

Given the large number of parameters involved in the TBNN, it is challenging to interpret the effect of the input features on its output. To get some additional insight on the impact of the different inputs, SHapley Additive exPlanations (SHAP) can be used. This approach, proposed by Lundberg and Lee (\citeyear{NIPS2017_7062}), aims to alleviate the "black box" limitations of neural networks through game theory. The SHAP values assign to each input feature a change in the expected model prediction when that feature is known, such that the added SHAP values of all features is equal to the model prediction. The SHAP Python library (Lundberg and Lee \citeyear{NIPS2017_7062}), is used to estimate the SHAP values for the TBNN model using a sampling technique. For visualization purposes, SHAP values were computed for a random selection of 10,000 data points from the model trained on the entire dataset, as shown in Figure \ref{fig:shapbxx}. Only the SHAP values for $b_{xx}^{\Delta}$ are displayed, as the values for other components are proportional to these. The 10 features with the highest mean absolute SHAP values are presented in descending order of impact. Interestingly, for the final output, $\boldsymbol{b}^{\Delta}$, the top three most important features are not dependent on MHD. This indicates that much of the physics captured by the model is relevant to non-MHD flows as well. Of the 10 features with the highest mean absolute SHAP values, only two, $q_{ASm}$ and $t_{turb} /t_{mag}$, are MHD-related features. Both of these features are approximately proportional to each other in the available MHD cases. They have high values near $z = 0$ and lower values everywhere else, therefore allowing the model to identify the region where MHD related effects are stronger. The contour plots of these features can be found in Appendix \ref{secA2}.

\subsubsection{Production Correction}

The results obtained using the SBNN models are summarised in Table \ref{tab:sbnn}. The RMSE values in this case are normalised by the total Root Mean Squared (RMS) of the correction field.

\begin{table}[b]
\centering
\caption{RMSE of $R$  of the SBNN models divided by the RMS of the target correction field $R$ for each CFD case. }
\label{tab:sbnn}
\begin{tabular}{@{}lllll@{}}
\cmidrule(l){2-5}
                                           & \multicolumn{4}{l}{RMSE($R$) / RMS($R$) {[}-{]}}                                                                                \\ \midrule
$R$ Model & $\mathrm{Ha}$ = 0 & $\mathrm{Ha}$ = 40 & $\mathrm{Ha}$ = 60 & $\mathrm{Ha}$ = 120 \\ \midrule
All Data Model       & 0.0871 & 0.282  & 0.328   & 0.828    \\ \midrule
Ha = 40 Test Model   & 0.0966 & 0.439   & 0.302   & 0.840    \\ \midrule
Ha = 60 Test Model  & 0.0952 & 0.287  & 0.421   & 0.831    \\ \midrule
Ha = 120 Test Model  & 0.0799 & 0.312   & 0.271   & 1.01     \\ \bottomrule
\end{tabular}
\end{table}

\begin{figure}[H]
 \centering
 \includegraphics[width=\textwidth]{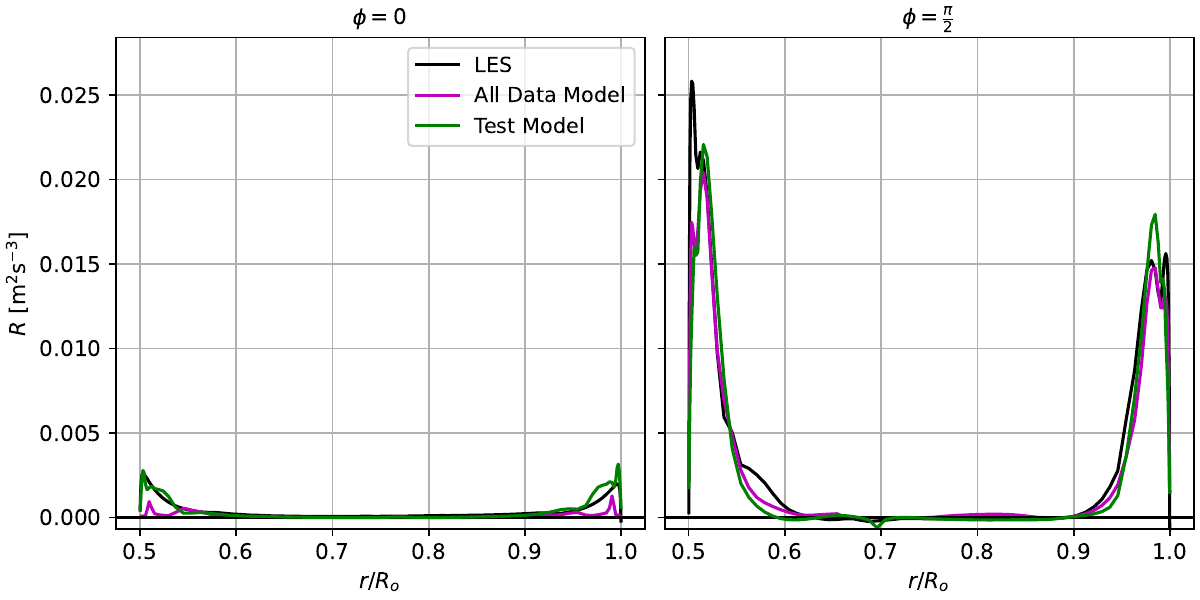}
 \caption{Profiles of $R$ on the $\mathrm{Ha}$ = 60 case for the SBNN All Data and test models.}
 \label{fig:pk60}
\end{figure}
\begin{figure}[H]
 \centering
 \includegraphics[width=\textwidth]{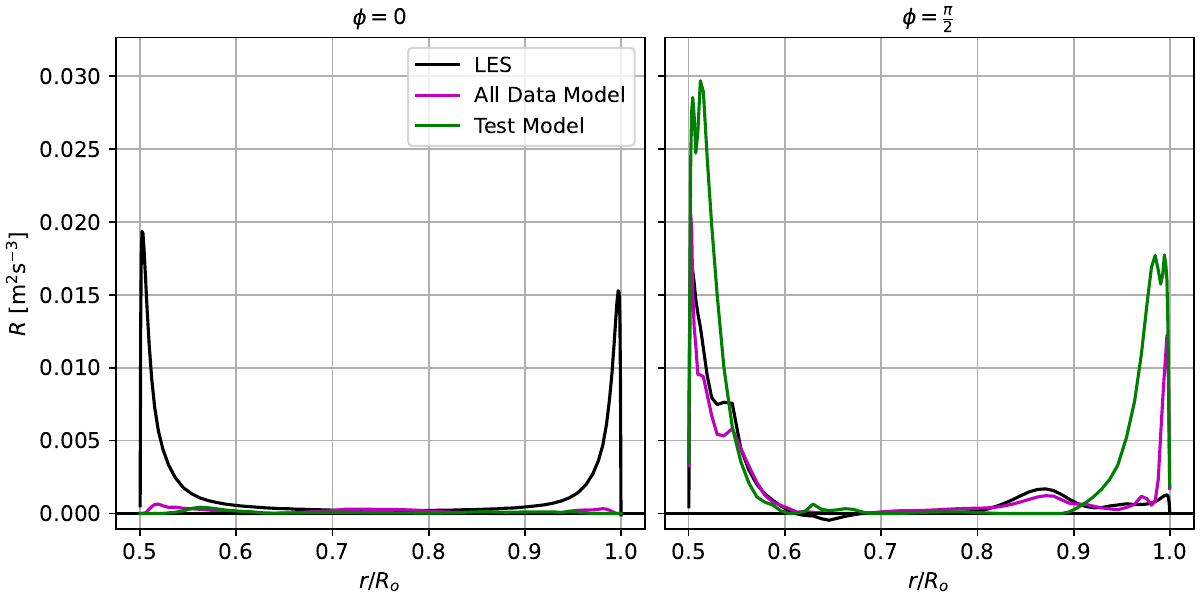}
 \caption{Profiles of $R$ on the $\mathrm{Ha}$ = 120 case for the SBNN All Data and test models.}
 \label{fig:pk120}
\end{figure}

Similarly to the TBNN, the predictions of the SBNN also worsen with increasing Hartmann number on a test case. The distribution of the correction field, its prediction by the SBNN and the resulting error are presented in Figure \ref{fig:pk60} for the $\mathrm{Ha}$ = 60 case. The profiles are plotted at two different azimuthal locations, see Figure \ref{fig:BCLES} for reference. The production correction is mostly present near the walls, being negligible in most of the domain. For the $\mathrm{Ha}$ = 60 case the predictions for both the All Data model and the test model match the target field reasonably. However, this is not the case for the $\mathrm{Ha}$ = 120 case, as the model cannot match the higher correction values at $\phi = 0$, see Figure \ref{fig:pk120}.

\begin{figure}[tbh]
 \centering
 \includegraphics[width=0.7\textwidth]{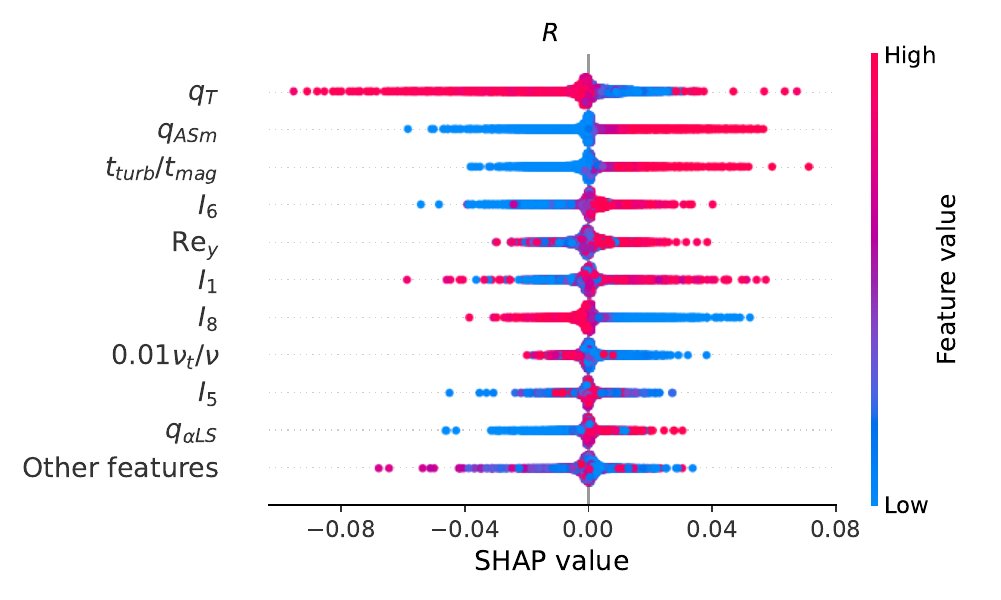}
 \caption{SHAP values for the input features of the SBNN with respect to $R$. The 10 features with the highest influence are shown separately in order.}
 \label{fig:shapPk}
\end{figure}

Based on the SHAP analysis, for the $R$ model the input features that are directly dependent on the Lorentz force also show a significant influence on the final output. In particular, three MHD features are found among the 10 most influential: $q_{ASm}$, $t_{turb}/ t_{mag}$ and $q_{\alpha LS}$, see Figure \ref{fig:shapPk}.   Of these, $q_{ASm}$ and $t_{turb}/ t_{mag}$ were also shown to be relevant for the $\boldsymbol{b}^{\Delta}$ model discussed in the previous Section, but not $q_{\alpha LS}$. A characteristic of the latter is that its value changes on both the inner and outer walls in the azimuthal direction of the annular flow cases, while the other MHD input features of the SBNN are approximately zero on the walls and show no significant variation in the azimuthal direction near the walls, see Figure \ref{fig:3MHDfeatures}. This makes it useful for modelling $R$, as most of the variation for this correction occurs in the inner boundary layer. The contour plots for these features can be found in Appendix \ref{secA2}. Both $I_1$ and $I_5$ (see Equations (\ref{eq:AI1}) and (\ref{eq:AI5}) respectively) remain among the most influential features, highlighting the significance of lower-order invariants in data-driven models based on the eddy viscosity hypothesis. Another observation is that, rather than the turbulence Reynolds number, the wall-distance-based Reynolds number, $\mathrm{Re}_y$, has a greater impact on modelling $R$. However, the fields of these local Reynolds numbers are approximately proportional for the flows under investigation. Hence, the SHAP analysis indicates that the data-driven models rely significantly on $\mathrm{Re}_y$ or $\mathrm{Re}_t$. This is because these parameters help the model distinguish between the different layers within the boundary layer, as identified by Jiang et al. (\citeyear{Jiang2021}) in their discussion of the "Non-Unique Mapping problem" of the generalised eddy hypothesis.

\subsection{\textit{A posteriori} Testing}

The outcome of the \textit{a posteriori} testing is summarised in Table \ref{tab:apost} in terms of the RMSE of the axial velocity $U_x$.
\begin{table}[h]
\centering
\caption{Summary table of the propagation results of all the MHD cases.}
\begin{tabular}{@{}llll@{}}
\cmidrule(l){2-4}
                       & \multicolumn{3}{l}{RMSE($U_x$)/ $U_b$} \\ \midrule\midrule
Turbulence Model       & $\mathrm{Ha}$ = 40     & $\mathrm{Ha}$ = 60    & $\mathrm{Ha}$ = 120    \\ \midrule\midrule
Baseline RANS      & 0.0577      & 0.0809     & 0.0707      \\ \midrule
Frozen RANS Propagation & 0.0120      & 0.0143     & 0.00888     \\ \midrule
All Data Model         & 0.0259      & 0.0254     & 0.0244      \\ \midrule
Test Models            & 0.0340       & 0.0479     & 0.0577      \\ \midrule\bottomrule
\end{tabular}
\label{tab:apost}
\end{table}
The tested models produce an improved mean velocity field compared to the RANS simulations using the baseline $k$-$\omega$ SST model. This also includes the test model for the $\mathrm{Ha}$ = 120 case, which is extrapolating to a higher Hartmann number. Unlike the $\boldsymbol{b}^{\Delta}$ and $R$ fields in \textit{a priori} testing, the $U_x$ fields of the All Data model simulations do not appear to decrease in accuracy for the higher Hartmann cases, while the test models still do. The explanation for this behaviour in the \textit{a posteriori} tests is that as the Re/Ha ratio decreases, the electromagnetic forces increase relative to the inertial forces. Hence, the effect of the Reynolds stress model on the mean velocity field is relatively lower due to the increased influence of the Lorentz force.

\begin{figure}[tbh]
\centering
 \includegraphics[width=\textwidth]{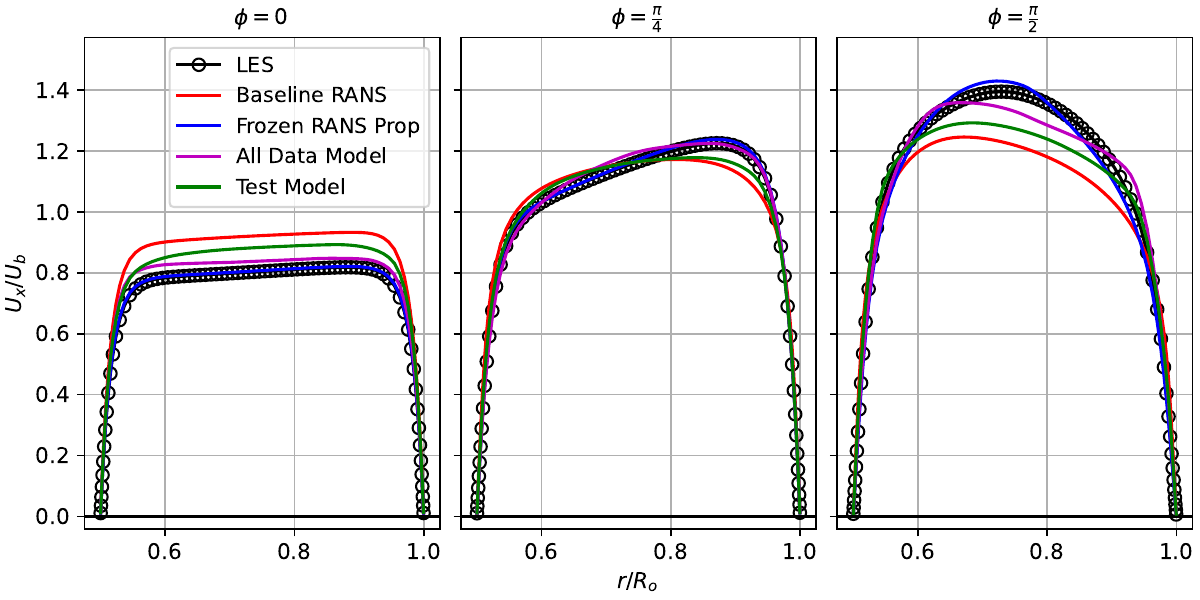}
 \caption{Velocity profiles of the $\mathrm{Ha}$ = 60 annular flow simulations with different turbulence models, including the TBNN and SBNN All Data and test models.}
 \label{fig:UxHa60}
\end{figure}

The results for the $\mathrm{Ha}$ = 60 case are further analysed next. Velocity profiles are shown in Figure \ref{fig:UxHa60} for three azimuthal locations in the pipe. The trends shown are also representative of the $\mathrm{Ha}$ = 40 case (not shown for conciseness). The Frozen RANS Propagation data shows the result if the regression error of the SBNN and TBNN were zero. This is done by using the correction fields that are extracted from the high fidelity data through Frozen RANS simulations explained in Section \ref{subsec:frozenRANS}. As expected based on the summary in Table \ref{tab:apost}, the All Data model achieves a significant improvement throughout the domain, while the test model exhibits values intermediate between the baseline and the All Data model profiles.

\begin{figure}[tbh]
\centering
 \includegraphics[width=\textwidth]{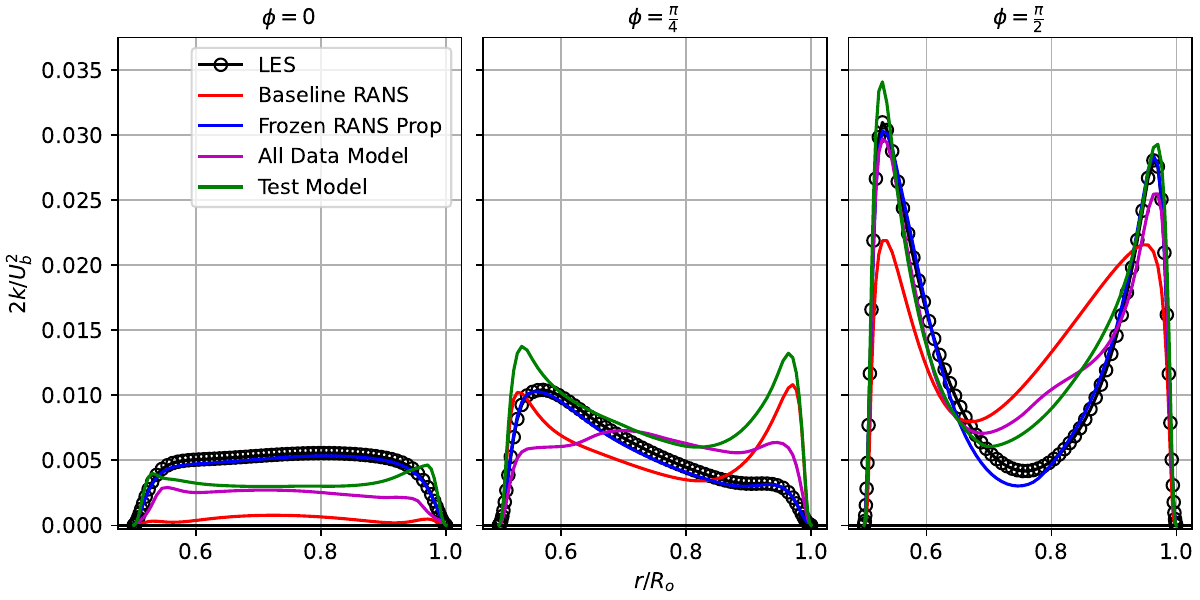}
 \caption{Turbulence kinetic energy of the $\mathrm{Ha}$ = 60 annular flow simulations with different turbulence models, including the TBNN and SBNN All Data and test models.}
 \label{fig:kHa60}
\end{figure}

The model should not only predict the turbulence anisotropy correctly, but also the turbulence intensity. Hence, we analyse the turbulence kinetic energy, $k$, as shown in Figure \ref{fig:kHa60}. It can be observed that the All Data model does not clearly outperform the test model in its prediction of $k$ throughout the domain. Since $P_k$ is also a function of $\boldsymbol{b}^{\Delta}$, the combined turbulence production of $P_k$ and $R$ depends also on $\boldsymbol{b}^{\Delta}$, see Equation (\ref{eq:pkfull}). This means that, as is the case in Figure \ref{fig:kHa60}, errors in $\boldsymbol{b}^{\Delta}$ can offset errors in $R$, resulting in a more accurate combined production $P_k + R$. The largest improvement relative to the baseline, occurs at $\phi = 0$. In all three MHD cases, $k$ in the baseline RANS is practically zero at this location with stronger Lorentz forces, which is not the case in the LES. The data-driven model provides a considerable improvement in this area. The overall error can be quantified using $\operatorname{RMSE}(k) / \operatorname{RMS}(k_{LES})$ as a metric, similarly to the metric used for $R$ in Table \ref{tab:sbnn}. Using this metric it can be shown that the MHD ML Turbulence models achieve a substantial improvement in the prediction, as the baseline RANS with the $k$-$\omega$ SST model has a $42 \%$ error, which the $\mathrm{Ha}$ = 60 test model improves to $33\%$ and the All Data model to $27 \%$ in the $\mathrm{Ha}$ = 60 case. 

\begin{figure}[tbh]
 \centering
 \includegraphics[width=\textwidth]{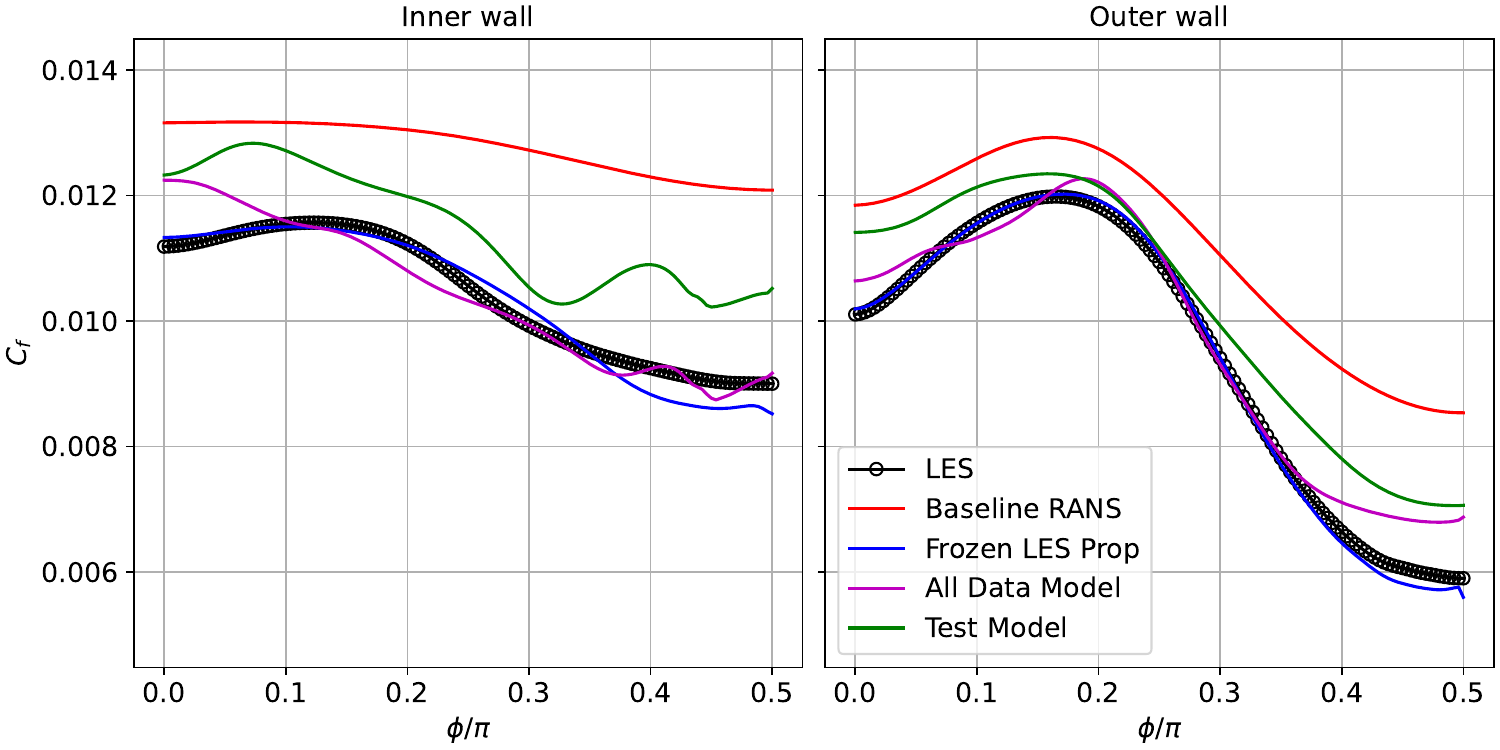}
 \caption{Friction coefficient $C_f$ along the outer and inner walls of the duct for the $\mathrm{Ha}$ = 60 case.}
 \label{fig:cfHa60}
\end{figure}

The pressure losses on the pipe depend on the skin friction coefficient $C_f$ along the walls. Therefore, it is important that the estimation of $C_f$ also improves when using the MHD ML turbulence model. This data is shown in Figure \ref{fig:cfHa60} for the $\mathrm{Ha}$ = 60 case. For this quantity, the All Data model demonstrates a high level of agreement with the results from the LES, effectively replicating its behavior, with a $\operatorname{RMSE}(C_f) / \operatorname{RMS}(C_{f, LES})$ of 4$\%$ compared to $20 \%$ for the baseline RANS. In contrast, the test model exhibits intermediate performance, falling between the baseline RANS and the LES predictions with a $\operatorname{RMSE} / \operatorname{RMS}(C_{f, LES})$ of $9 \%$. Furthermore, the skin friction coefficient of the simulations using the data-driven model show an artificial behaviour in the form of oscillations on the inner wall that are not present in the LES data. This additional variability can be quantified through the standard deviation of the first derivative of $C_f$ on the azimuthal direction $\frac{\partial C_f}{\partial \phi}$ which increases by 72$\%$ for the $\mathrm{Ha}$ = 60 test model with respect to the LES on the inner wall. This indicates a lack of viscous diffusion in this region which for this case did not result in numerical instabilities, but should be monitored when testing the model in future study cases.


\begin{figure}[tbh]
\centering
 \includegraphics[width=\textwidth]{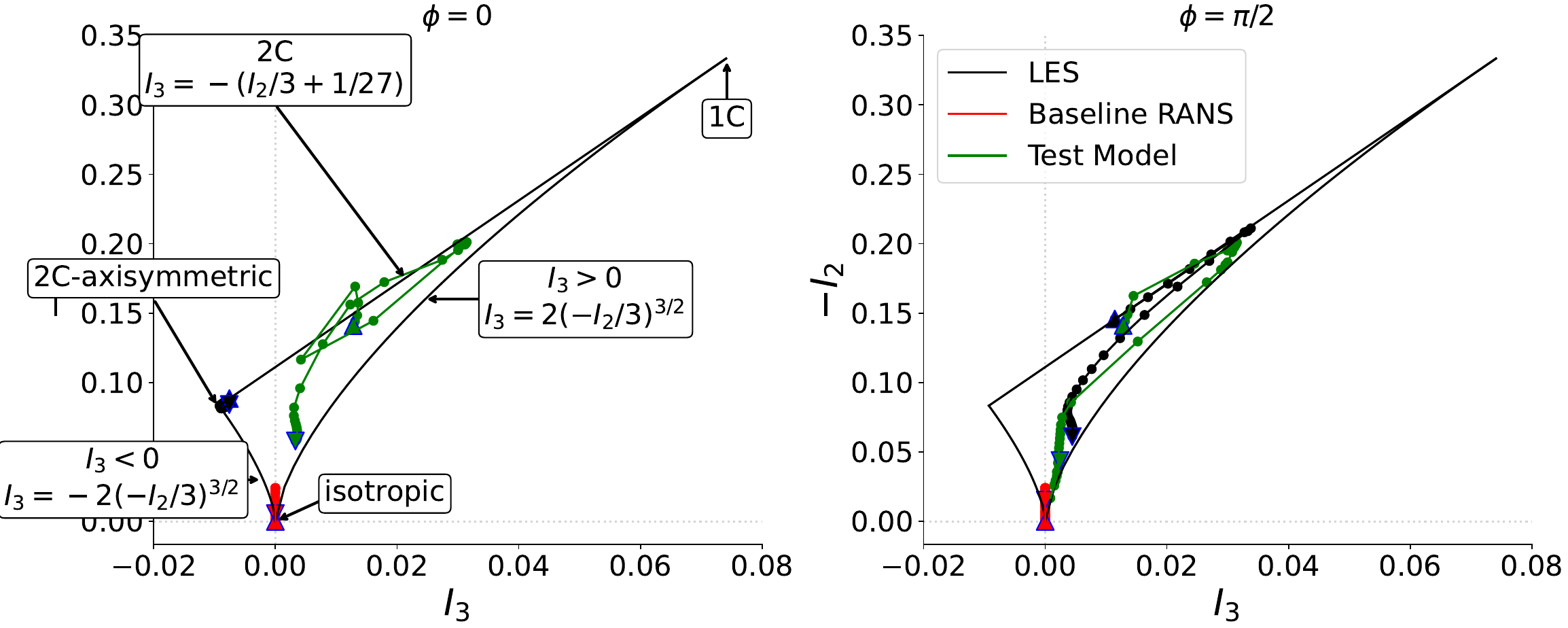}
 \caption{Lumley triangle plot for the $\mathrm{Ha}$ = 60 case. The invariant values at the cell centers are plotted from the inner wall to the center line. The triangle marker pointing upwards corresponds to the cell closest to the wall.}
 \label{fig:Lumley60}
\end{figure}

Finally, to visualise the turbulence anisotropy more clearly the Lumley triangle plot is used (Lumley \citeyear{lumley1979computational}). In this case the invariants of $\boldsymbol{b}$ for the $\mathrm{Ha}$ = 60 case are shown at the $\phi = 0$ and $\phi = \pi / 2 $ locations. These are presented in Figure \ref{fig:Lumley60}. This technique consists in mapping the second and third invariants of the turbulence anisotropy $\boldsymbol{b}$, where $I_2 = I_3 = 0$ represents the fully isotropic state, the top right vertex represents the 1 component state, where $I_2 = -1/3$, $I_3 = 2/27$ and the top left vertex represents the two component axisymmetric state, where $I_2 = -1/12 $, $I_3 = -1/108$. The area within the triangle and on its edges represents intermediate states. Observing Figure \ref{fig:Lumley60}, a visible area of improvement is that the test model predictions do not always remain within the triangle, meaning that those values are not physically realizable. 

Overall, the data-driven model matches the anisotropy states of the LES results better than the baseline RANS, whose predictions are instead limited to the plain strain line at $I_3 = 0$. This is particularly noticeable at $\phi = \pi / 2$. However, at $\phi = 0$, where the MHD related effects are stronger, the data-driven model does not match the fully quasi-2D turbulence state predicted by the LES. This aligns with the observations about the TBNN predictions for $\boldsymbol{b}^{\Delta}$ in \textit{a priori} testing (see Section \ref{subsec:tbnnapriori}). At this location the results only approach the 2C axisymmetric expansion line that defines quasi-2D turbulence states. This is due to a higher error in the estimation of the normal stress components of the turbulence anisotropy correction, such as $b_{xx}^{\Delta}$, as shown in Figure \ref{fig:ha60_full}. Since for incompressible flows $I_1 = b_{xx} + b_{yy} + b_{zz} = 0$, this also implies an error in $b_{yy}^{\Delta}$ and $b_{zz}^{\Delta}$. 


\section{Conclusion and Recommendations}
\label{sec:conc}

This work introduces a data-driven turbulence modelling framework for MHD flows in annular pipes. $k$-corrective Frozen RANS simulations are used to obtain the target correction fields for the turbulence anisotropy tensor $\boldsymbol{b}^{\Delta}$ and the turbulence production $R$. To obtain a Galilean and frame invariant model, the generalised eddy hypothesis is used to obtain the input features for the data-driven model. To capture MHD effects, the set of tensors used to generate the invariant basis is extended with the $\boldsymbol{A_L}$ and $\boldsymbol{A_k}$ antisymmetric tensors, where $\boldsymbol{A_L}$ is a tensor representation of the Lorentz force vector. For the regression of the anisotropy correction, the TBNN is used, while for the production correction the SBNN is proposed as a solution.

To reduce model complexity, higher-order tensors providing redundant information are discarded using orthogonal projections. Furthermore,  to minimise the complexity of the ${\tilde g}_n$ basis coefficient functions, and keep their values bounded to O(1), a different set of non dimensionalisation factors is proposed for computing the basis tensors than that used for the invariants. The model was implemented in a RANS solver using an open loop approach, where the TBNN and SBNN are only queried one time, instead of every iteration.

The \textit{a priori} results show that the model can partly capture the turbulence anisotropy trends in MHD annular flow cases. However, its performance deteriorates in cases with higher Hartmann numbers, and quasi-2D turbulence states are not predicted accurately. In contrast, the \textit{a posteriori} tests demonstrate significant improvements in the predicted mean flow velocity field, turbulent kinetic energy field, and friction coefficient compared to the baseline RANS results obtained with the $k$-$\omega$ SST model. The reduction in error was the smallest for the extrapolation test, where the model was trained in $\mathrm{Ha}$ = 0, 40 and 60 cases and tested on the $\mathrm{Ha}$ = 120 case.

For future developments of this methodology, certain improvements could provide a substantial reduction in the \textit{a priori} and \textit{a posteriori} errors of the data-driven model. One key improvement would be implementing the model in a closed-loop framework, where it is queried multiple times during the simulation. This would allow the model to be trained using features extracted from LES data instead of RANS, leading to more physically accurate coefficients and improved accuracy, particularly when the baseline RANS solution deviates significantly from the LES mean flow. Another potential improvement could be training two separate models: one for the inner boundary layer and another for the outer boundary layer and freestream. These two models could then be blended using a function based on wall distance, similarly to $k$-$\omega$ SST.

\section*{Acknowledgements}
The authors further gratefully acknowledge Dr. Hao Xia from Loughborough University for the LES data provided and useful comments.


\section*{Declarations}

\textbf{Funding Information} \\
Francesco Fico and Ivan Langella acknowledge the funding support from Engineering and Physical Science Research Council (EPSRC) under Grant No. EP/R513088/1 and UK Atomic Energy Authority (UKAEA).








\begin{appendices}




\section{List of Invariants and Basis Tensors}\label{secA2}

The original 10 basis tensors of the generalised eddy hypothesis are (Pope \citeyear{Pope1975})

\begin{equation}
\label{eq:tensorspope}
\left.\begin{array}{ll}
\boldsymbol{T}^{(1)}=\hat{\boldsymbol{S}} & \boldsymbol{T}^{(6)}=\hat{\boldsymbol{\Omega}}^2 \hat{\boldsymbol{S}}+\boldsymbol{S}^2-\frac{2}{3} \boldsymbol{I} \cdot \operatorname{Tr}\left(\hat{\boldsymbol{S}} \hat{\boldsymbol{\Omega}}^2\right) \\
\boldsymbol{T}^{(2)}=\hat{\boldsymbol{S}} \hat{\boldsymbol{\Omega}}-\hat{\boldsymbol{\Omega}} \hat{\boldsymbol{S}} & \boldsymbol{T}^{(7)}=\hat{\boldsymbol{\Omega}} \hat{\boldsymbol{S}} \hat{\boldsymbol{\Omega}}^2-\hat{\boldsymbol{\Omega}}^2 \hat{\boldsymbol{S}}\hat{\boldsymbol{\Omega}} \\
\boldsymbol{T}^{(3)}=\hat{\boldsymbol{S}}^2-\frac{1}{3} \boldsymbol{I} \cdot \operatorname{Tr}\left(\hat{\boldsymbol{S}}^2\right) & \boldsymbol{T}^{(8)}=\hat{\boldsymbol{S}} \hat{\boldsymbol{\Omega}} \hat{\boldsymbol{S}}^2-\hat{\boldsymbol{S}}^2 \hat{\boldsymbol{\Omega}} \hat{\boldsymbol{S}} \\
\boldsymbol{T}^{(4)}=\hat{\boldsymbol{\Omega}}^2-\frac{1}{3} \boldsymbol{I} \cdot \operatorname{Tr}\left(\hat{\boldsymbol{\Omega}}^2\right) & \boldsymbol{T}^{(9)}=\hat{\boldsymbol{\Omega}}^2 \hat{\boldsymbol{S}}^2+\hat{\boldsymbol{S}}^2 \hat{\boldsymbol{\Omega}}^2-\frac{2}{3} \boldsymbol{I} \cdot \operatorname{Tr}\left(\hat{\boldsymbol{S}}^2 \hat{\boldsymbol{\Omega}}^2\right) \\
\boldsymbol{T}^{(5)}=\hat{\boldsymbol{\Omega}} \hat{\boldsymbol{S}}^2-\hat{\boldsymbol{S}}^2 \hat{\boldsymbol{\Omega}} & \boldsymbol{T}^{(10)}=\hat{\boldsymbol{\Omega}} \hat{\boldsymbol{S}}^2 \hat{\boldsymbol{\Omega}}^2-\hat{\boldsymbol{\Omega}}^2 \hat{\boldsymbol{S}}^2 \hat{\boldsymbol{\Omega}}
\end{array}\right\}
\end{equation}

Below is presented the complete list of the 47 invariants that can be obtained from the following set of non-dimensional antisymmetric and symmetric tensors using the Cayley-Hamilton theorem: $\hat{\boldsymbol{S}}$, $\hat{\boldsymbol{\Omega}}$, $\hat{\boldsymbol{A_K}}$ and $\hat{\boldsymbol{A_L}}$. The list was derived using the methodology proposed by Wu et al. (\citeyear{Wu2018}), who used a tensor based on the pressure gradient in place of $\hat{\boldsymbol{A_L}}$. 

\noindent
\begin{minipage}{.49\linewidth}
\begin{equation}
\label{eq:AI1}
    I_1 = \operatorname{tr}(\hat{\boldsymbol{S}}^2)
\end{equation}
\end{minipage}
\begin{minipage}{.49\linewidth}
\begin{equation}
    I_2 = \operatorname{tr}(\hat{\boldsymbol{\Omega}}^2)
\end{equation}
\end{minipage}
\noindent
\begin{minipage}{.49\linewidth}
\begin{equation}
    I_3 = \operatorname{tr}(\hat{\boldsymbol{S}}^3)
\end{equation}
\end{minipage}
\begin{minipage}{.49\linewidth}
\begin{equation}
    I_4 = \operatorname{tr}(\hat{\boldsymbol{\Omega}}^2 \hat{\boldsymbol{S}})
\end{equation}
\end{minipage}
\noindent
\begin{minipage}{.49\linewidth}
\begin{equation}
\label{eq:AI5}
    I_5 = \operatorname{tr}(\hat{\boldsymbol{\Omega}}^2 \hat{\boldsymbol{S}}^2)
\end{equation}
\end{minipage}
\begin{minipage}{.49\linewidth}
\begin{equation}
    I_6 = \operatorname{tr}(\hat{\boldsymbol{A_k}}^2)
\end{equation}
\end{minipage}
\noindent
\begin{minipage}{.49\linewidth}
\begin{equation}
    I_7 = \operatorname{tr}(\hat{\boldsymbol{A_k}}^2 \hat{\boldsymbol{S}})
\end{equation}
\end{minipage}
\begin{minipage}{.49\linewidth}
\begin{equation}
    I_8 = \operatorname{tr}(\hat{\boldsymbol{A_k}}^2 \hat{\boldsymbol{S}}^2)
\end{equation}
\end{minipage}
\noindent
\begin{minipage}{.49\linewidth}
\begin{equation}
    I_9 = \operatorname{tr}(\hat{\boldsymbol{\Omega}} \hat{\boldsymbol{A_k}})
\end{equation}
\end{minipage}
\begin{minipage}{.49\linewidth}
\begin{equation}
    I_{10} = \operatorname{tr}(\hat{\boldsymbol{\Omega}} \hat{\boldsymbol{A_k}} \hat{\boldsymbol{S}})
\end{equation}
\end{minipage}
\noindent
\begin{minipage}{.49\linewidth}
\begin{equation}
    I_{11} = \operatorname{tr}(\hat{\boldsymbol{A_k}}^2 \hat{\boldsymbol{S}}
    \hat{\boldsymbol{A_k}} \hat{\boldsymbol{S}}^2)
\end{equation}
\end{minipage}
\begin{minipage}{.49\linewidth}
\begin{equation}
    I_{12} = \operatorname{tr}(\hat{\boldsymbol{\Omega}} \hat{\boldsymbol{A_k}} \hat{\boldsymbol{S}}^2)
\end{equation}
\end{minipage}
\noindent
\begin{minipage}{.49\linewidth}
\begin{equation}
    I_{13} = \operatorname{tr}(\hat{\boldsymbol{\Omega}}^2 \hat{\boldsymbol{A_k}} \hat{\boldsymbol{S}})
\end{equation}
\end{minipage}
\begin{minipage}{.49\linewidth}
\begin{equation}
    I_{14} = \operatorname{tr}(\hat{\boldsymbol{\Omega}}^2 \hat{\boldsymbol{A_k}} \hat{\boldsymbol{S}}^2)
\end{equation}
\end{minipage}
\noindent
\begin{minipage}{.49\linewidth}
\begin{equation}
    I_{15} = \operatorname{tr}(\hat{\boldsymbol{A_k}}^2 \hat{\boldsymbol{\Omega}} \hat{\boldsymbol{S}})
\end{equation}
\end{minipage}
\begin{minipage}{.49\linewidth}
\begin{equation}
    I_{16} = \operatorname{tr}(\hat{\boldsymbol{A_k}}^2 \hat{\boldsymbol{\Omega}} \hat{\boldsymbol{S}}^2)
\end{equation}
\end{minipage}
\noindent
\begin{minipage}{.49\linewidth}
\begin{equation}
    I_{17} = \operatorname{tr}(\hat{\boldsymbol{\Omega}}^2 \hat{\boldsymbol{S}}
    \hat{\boldsymbol{A_k}} \hat{\boldsymbol{S}}^2)
\end{equation}
\end{minipage}
\begin{minipage}{.49\linewidth}
\begin{equation}
    I_{18} = \operatorname{tr}(\hat{\boldsymbol{A_k}}^2 \hat{\boldsymbol{S}}
    \hat{\boldsymbol{\Omega}} \hat{\boldsymbol{S}}^2)
\end{equation}
\end{minipage}
\noindent
\begin{minipage}{.49\linewidth}
\begin{equation}
    I_{19} = \operatorname{tr}(\hat{\boldsymbol{A_L}}^2)
\end{equation}
\end{minipage}
\begin{minipage}{.49\linewidth}
\begin{equation}
    I{20} = \operatorname{tr}(\hat{\boldsymbol{A_L}}^2 \hat{\boldsymbol{S}})
\end{equation}
\end{minipage}
\noindent
\begin{minipage}{.49\linewidth}
\begin{equation}
    I_{21} = \operatorname{tr}(\hat{\boldsymbol{A_L}}^2 \hat{\boldsymbol{S}}^2)
\end{equation}
\end{minipage}
\begin{minipage}{.49\linewidth}
\begin{equation}
    I_{22} = \operatorname{tr}(\hat{\boldsymbol{\Omega}} \hat{\boldsymbol{A_L}})
\end{equation}
\end{minipage}
\noindent
\begin{minipage}{.49\linewidth}
\begin{equation}
    I_{23} = \operatorname{tr}(\hat{\boldsymbol{\Omega}} \hat{\boldsymbol{A_L}} \hat{\boldsymbol{S}})
\end{equation}
\end{minipage}
\begin{minipage}{.49\linewidth}
\begin{equation}
    I_{24} = \operatorname{tr}(\hat{\boldsymbol{A_L}}^2 \hat{\boldsymbol{S}}
    \hat{\boldsymbol{A_L}} \hat{\boldsymbol{S}}^2)
\end{equation}
\end{minipage}
\noindent
\begin{minipage}{.49\linewidth}
\begin{equation}
    I_{25} = \operatorname{tr}(\hat{\boldsymbol{\Omega}} \hat{\boldsymbol{A_L}} \hat{\boldsymbol{S}}^2)
\end{equation}
\end{minipage}
\begin{minipage}{.49\linewidth}
\begin{equation}
    I_{26} = \operatorname{tr}(\hat{\boldsymbol{\Omega}}^2 \hat{\boldsymbol{A_L}} \hat{\boldsymbol{S}})
\end{equation}
\end{minipage}
\noindent
\begin{minipage}{.49\linewidth}
\begin{equation}
    I_{27} = \operatorname{tr}(\hat{\boldsymbol{\Omega}}^2 \hat{\boldsymbol{A_L}} \hat{\boldsymbol{S}}^2)
\end{equation}
\end{minipage}
\begin{minipage}{.49\linewidth}
\begin{equation}
    I_{28} = \operatorname{tr}(\hat{\boldsymbol{A_L}}^2 \hat{\boldsymbol{\Omega}} \hat{\boldsymbol{S}})
\end{equation}
\end{minipage}
\noindent
\begin{minipage}{.49\linewidth}
\begin{equation}
    I_{29} = \operatorname{tr}(\hat{\boldsymbol{A_L}}^2 \hat{\boldsymbol{\Omega}} \hat{\boldsymbol{S}}^2)
\end{equation}
\end{minipage}
\begin{minipage}{.49\linewidth}
\begin{equation}
    I_{30} = \operatorname{tr}(\hat{\boldsymbol{\Omega}}^2 \hat{\boldsymbol{S}}
    \hat{\boldsymbol{A_L}} \hat{\boldsymbol{S}}^2)
\end{equation}
\end{minipage}
\noindent
\begin{minipage}{.49\linewidth}
\begin{equation}
    I_{31} = \operatorname{tr}(\hat{\boldsymbol{A_L}}^2 \hat{\boldsymbol{S}}
    \hat{\boldsymbol{\Omega}} \hat{\boldsymbol{S}}^2)
\end{equation}
\end{minipage}
\begin{minipage}{.49\linewidth}
\begin{equation}
    I_{32} = \operatorname{tr}(\hat{\boldsymbol{A_L}}
    \hat{\boldsymbol{A_k}} \hat{\boldsymbol{S}})
\end{equation}
\end{minipage}
\noindent
\begin{minipage}{.49\linewidth}
\begin{equation}
    I_{33} = \operatorname{tr}(\hat{\boldsymbol{A_L}}
    \hat{\boldsymbol{A_k}} \hat{\boldsymbol{S}}^2)
\end{equation}
\end{minipage}
\begin{minipage}{.49\linewidth}
\begin{equation}
    I_{34} = \operatorname{tr}(\hat{\boldsymbol{A_k}}^2
    \hat{\boldsymbol{A_L}} \hat{\boldsymbol{S}})
\end{equation}
\end{minipage}
\noindent
\begin{minipage}{.49\linewidth}
\begin{equation}
    I_{35} = \operatorname{tr}(\hat{\boldsymbol{A_L}}^2
    \hat{\boldsymbol{A_k}} \hat{\boldsymbol{S}})
\end{equation}
\end{minipage}
\begin{minipage}{.49\linewidth}
\begin{equation}
    I_{36} = \operatorname{tr}(\hat{\boldsymbol{A_k}}^2
    \hat{\boldsymbol{A_L}} \hat{\boldsymbol{S}}^2)
\end{equation}
\end{minipage}
\noindent
\begin{minipage}{.49\linewidth}
\begin{equation}
    I_{37} = \operatorname{tr}(\hat{\boldsymbol{A_L}}^2
    \hat{\boldsymbol{A_k}} \hat{\boldsymbol{S}}^2)
\end{equation}
\end{minipage}
\begin{minipage}{.49\linewidth}
\begin{equation}
    I_{38} = \operatorname{tr}(\hat{\boldsymbol{A_k}}^2
    \hat{\boldsymbol{S}}
    \hat{\boldsymbol{A_L}} \hat{\boldsymbol{S}}^2)
\end{equation}
\end{minipage}
\noindent
\begin{minipage}{.49\linewidth}
\begin{equation}
    I_{39} = \operatorname{tr}(\hat{\boldsymbol{A_L}}^2
    \hat{\boldsymbol{S}}
    \hat{\boldsymbol{A_k}} \hat{\boldsymbol{S}}^2)
\end{equation}
\end{minipage}
\begin{minipage}{.49\linewidth}
\begin{equation}
    I_{40} = \operatorname{tr}(\hat{\boldsymbol{\Omega}}
    \hat{\boldsymbol{A_L}} \hat{\boldsymbol{A_k}})
\end{equation}
\end{minipage}
\noindent
\begin{minipage}{.49\linewidth}
\begin{equation}
    I_{41} = \operatorname{tr}(\hat{\boldsymbol{\Omega}}
    \hat{\boldsymbol{A_L}} \hat{\boldsymbol{A_k}} \hat{\boldsymbol{S}})
\end{equation}
\end{minipage}
\begin{minipage}{.49\linewidth}
\begin{equation}
    I_{42} = \operatorname{tr}(\hat{\boldsymbol{\Omega}}
    \hat{\boldsymbol{A_k}} \hat{\boldsymbol{A_L}} \hat{\boldsymbol{S}})
\end{equation}
\end{minipage}
\noindent
\begin{minipage}{.49\linewidth}
\begin{equation}
    I_{43} = \operatorname{tr}(\hat{\boldsymbol{\Omega}}
    \hat{\boldsymbol{A_L}} \hat{\boldsymbol{A_k}} \hat{\boldsymbol{S}}^2)
\end{equation}
\end{minipage}
\begin{minipage}{.49\linewidth}
\begin{equation}
    I_{44} = \operatorname{tr}(\hat{\boldsymbol{\Omega}}
    \hat{\boldsymbol{A_k}} \hat{\boldsymbol{A_L}} \hat{\boldsymbol{S}}^2)
\end{equation}
\end{minipage}
\noindent
\begin{minipage}{.49\linewidth}
\begin{equation}
     I_{45} = \operatorname{tr}(\hat{\boldsymbol{\Omega}}
    \hat{\boldsymbol{A_L}}
    \hat{\boldsymbol{S}}
    \hat{\boldsymbol{A_k}} \hat{\boldsymbol{S}}^2)
\end{equation}
\end{minipage}
\begin{minipage}{.49\linewidth}
\begin{equation}
    I_{46} = \operatorname{tr}(\hat{\boldsymbol{\Omega}}^2 \hat{\boldsymbol{S}}
    \hat{\boldsymbol{\Omega}} \hat{\boldsymbol{S}}^2)
\end{equation}
\end{minipage}
\noindent
\begin{minipage}{.49\linewidth}
\begin{equation}
    I_{47} = \operatorname{tr}(\hat{\boldsymbol{A_L}} \hat{\boldsymbol{A_k}})
\end{equation}
\end{minipage}

\vspace{5mm}

\begin{figure}[h!]
 \centering
 \includegraphics[width=0.7\textwidth]{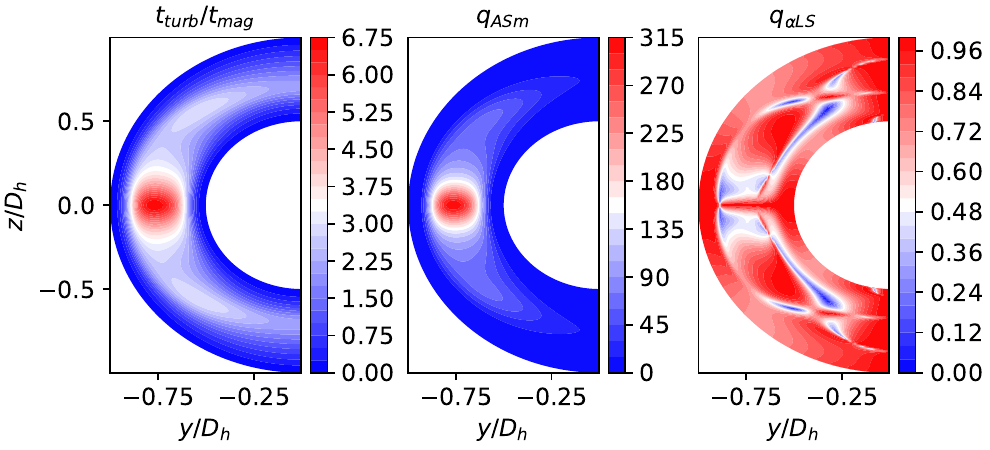}
 \caption{Contour plots of the 3 MHD-based features with the most impact on the closure model's predictions for the $\mathrm{Ha}$ = 60 case.}
 \label{fig:3MHDfeatures}
\end{figure}

\noindent
The contour plots of the MHD features with the most impact on the closure model's predictions according to the SHAP analysis are shown in Figure \ref{fig:3MHDfeatures}.

Finally, for the TBNN and SBNN models the full invariant basis was not used, but instead a subset of the invariant basis combined with a  subset of the additional features presented in Table \ref{tab:newfeatures}. The lists of used features are 

\begin{equation}
\label{eq:listtbnn}
\begin{aligned}
    \boldsymbol{q}_{TBNN} = & [I_1, I_5,I_6, I_8, I_9, I_{11}, I_{12}, I_{15}, I_{18}, I_{19}, I_{21}, I_{32}, I_{35},\\
    & I_{36}, I_{38}, I_{43}, \operatorname{Re}_t , t_{turb} / t_{mag} , \nu_t / 100 \nu, q_T, t_{mean} / t_{mag}, q _{AS m}] ,
\end{aligned}
\end{equation}

\begin{equation}
\label{eq:listsbnn}
\begin{aligned}
    \boldsymbol{q}_{SBNN} = & [I_1, I_5,I_6, I_8,I_9, I_{11}, I_{12}, I_{15}, I_{19}, I_{20}, I_{21}, I_{32}, I_{35},\\
    & I_{40}, I_{43}, I_{44}, \operatorname{Re}_t , t_{turb} / t_{mag} , \operatorname{Re}_y, \nu_t / 100 \nu, q_T, t_{mean} / t_{mag},\\
    & q _{AS \omega}, q _{AS m}, q _{A}] .
\end{aligned}
\end{equation}




\end{appendices}


\bibliography{sn-bibliography}


\begin{thebibliography}{31}
\ifx \bisbn   \undefined \def \bisbn  #1{ISBN #1}\fi
\ifx \binits  \undefined \def \binits#1{#1}\fi
\ifx \bauthor  \undefined \def \bauthor#1{#1}\fi
\ifx \batitle  \undefined \def \batitle#1{#1}\fi
\ifx \bjtitle  \undefined \def \bjtitle#1{#1}\fi
\ifx \bvolume  \undefined \def \bvolume#1{\textbf{#1}}\fi
\ifx \byear  \undefined \def \byear#1{#1}\fi
\ifx \bissue  \undefined \def \bissue#1{#1}\fi
\ifx \bfpage  \undefined \def \bfpage#1{#1}\fi
\ifx \blpage  \undefined \def \blpage #1{#1}\fi
\ifx \burl  \undefined \def \burl#1{\textsf{#1}}\fi
\ifx \doiurl  \undefined \def \doiurl#1{\url{https://doi.org/#1}}\fi
\ifx \betal  \undefined \def \betal{\textit{et al.}}\fi
\ifx \binstitute  \undefined \def \binstitute#1{#1}\fi
\ifx \binstitutionaled  \undefined \def \binstitutionaled#1{#1}\fi
\ifx \bctitle  \undefined \def \bctitle#1{#1}\fi
\ifx \beditor  \undefined \def \beditor#1{#1}\fi
\ifx \bpublisher  \undefined \def \bpublisher#1{#1}\fi
\ifx \bbtitle  \undefined \def \bbtitle#1{#1}\fi
\ifx \bedition  \undefined \def \bedition#1{#1}\fi
\ifx \bseriesno  \undefined \def \bseriesno#1{#1}\fi
\ifx \blocation  \undefined \def \blocation#1{#1}\fi
\ifx \bsertitle  \undefined \def \bsertitle#1{#1}\fi
\ifx \bsnm \undefined \def \bsnm#1{#1}\fi
\ifx \bsuffix \undefined \def \bsuffix#1{#1}\fi
\ifx \bparticle \undefined \def \bparticle#1{#1}\fi
\ifx \barticle \undefined \def \barticle#1{#1}\fi
\bibcommenthead
\ifx \bconfdate \undefined \def \bconfdate #1{#1}\fi
\ifx \botherref \undefined \def \botherref #1{#1}\fi
\ifx \url \undefined \def \url#1{\textsf{#1}}\fi
\ifx \bchapter \undefined \def \bchapter#1{#1}\fi
\ifx \bbook \undefined \def \bbook#1{#1}\fi
\ifx \bcomment \undefined \def \bcomment#1{#1}\fi
\ifx \oauthor \undefined \def \oauthor#1{#1}\fi
\ifx \citeauthoryear \undefined \def \citeauthoryear#1{#1}\fi
\ifx \endbibitem  \undefined \def \endbibitem {}\fi
\ifx \bconflocation  \undefined \def \bconflocation#1{#1}\fi
\ifx \arxivurl  \undefined \def \arxivurl#1{\textsf{#1}}\fi
\csname PreBibitemsHook\endcsname

\bibitem[\protect\citeauthoryear{Cinnella}{2024}]{cinnella2024data}
\begin{botherref}
\oauthor{\bsnm{Cinnella}, \binits{P.}}:
Data-driven turbulence modeling.
arXiv preprint arXiv:2404.09074
(2024)
\end{botherref}
\endbibitem

\bibitem[\protect\citeauthoryear{Fico et~al.}{2023}]{Fico2023}
\begin{botherref}
\oauthor{\bsnm{Fico}, \binits{F.}},
\oauthor{\bsnm{Langella}, \binits{I.}},
\oauthor{\bsnm{Xia}, \binits{H.}}:
Large-eddy simulation of magnetohydrodynamics and heat transfer in annular pipe liquid metal flow.
Physics of Fluids
\textbf{35}
(2023)
\doiurl{10.1063/5.0143687}
\end{botherref}
\endbibitem

\bibitem[\protect\citeauthoryear{Huijing et~al.}{2021}]{huijing2021data}
\begin{barticle}
\bauthor{\bsnm{Huijing}, \binits{J.P.}},
\bauthor{\bsnm{Dwight}, \binits{R.P.}},
\bauthor{\bsnm{Schmelzer}, \binits{M.}}:
\batitle{Data-driven rans closures for three-dimensional flows around bluff bodies}.
\bjtitle{Computers \& Fluids}
\bvolume{225},
\bfpage{104997}
(\byear{2021})
\end{barticle}
\endbibitem

\bibitem[\protect\citeauthoryear{Jigjid et~al.}{2024}]{jigjid2024simple}
\begin{bchapter}
\bauthor{\bsnm{Jigjid}, \binits{K.}},
\bauthor{\bsnm{Dwight}, \binits{R.}},
\bauthor{\bsnm{Allaerts}, \binits{D.}},
\bauthor{\bsnm{Steiner}, \binits{J.}}:
\bctitle{A simple rans closure for wind-farms under neutral atmospheric conditions: Preliminary findings}.
In: \bbtitle{Journal of Physics: Conference Series},
vol. \bseriesno{2767},
p. \bfpage{092104}
(\byear{2024}).
\bcomment{IOP Publishing}
\end{bchapter}
\endbibitem

\bibitem[\protect\citeauthoryear{Ji and Gardner}{1997}]{Ji1997}
\begin{barticle}
\bauthor{\bsnm{Ji}, \binits{H.-C.}},
\bauthor{\bsnm{Gardner}, \binits{R.}}:
\batitle{Numerical analysis of turbulent pipe flow in a transverse magnetic field}.
\bjtitle{International journal of heat and mass transfer}
\bvolume{40}(\bissue{8}),
\bfpage{1839}--\blpage{1851}
(\byear{1997})
\end{barticle}
\endbibitem

\bibitem[\protect\citeauthoryear{Jiang et~al.}{2021}]{Jiang2021}
\begin{botherref}
\oauthor{\bsnm{Jiang}, \binits{C.}},
\oauthor{\bsnm{Vinuesa}, \binits{R.}},
\oauthor{\bsnm{Chen}, \binits{R.}},
\oauthor{\bsnm{Mi}, \binits{J.}},
\oauthor{\bsnm{Laima}, \binits{S.}},
\oauthor{\bsnm{Li}, \binits{H.}}:
An interpretable framework of data-driven turbulence modeling using deep neural networks.
Physics of Fluids
\textbf{33}
(2021)
\doiurl{10.1063/5.0048909}
\end{botherref}
\endbibitem

\bibitem[\protect\citeauthoryear{Kingma and Ba}{2014}]{kingma2014adam}
\begin{botherref}
\oauthor{\bsnm{Kingma}, \binits{D.P.}},
\oauthor{\bsnm{Ba}, \binits{J.}}:
Adam: A method for stochastic optimization.
arXiv preprint arXiv:1412.6980
(2014)
\end{botherref}
\endbibitem

\bibitem[\protect\citeauthoryear{Kaandorp and Dwight}{2020}]{Kaandorp2020}
\begin{botherref}
\oauthor{\bsnm{Kaandorp}, \binits{M.L.A.}},
\oauthor{\bsnm{Dwight}, \binits{R.P.}}:
Data-driven modelling of the reynolds stress tensor using random forests with invariance.
Computers and Fluids
\textbf{202}
(2020)
\doiurl{10.1016/j.compfluid.2020.104497}
\end{botherref}
\endbibitem

\bibitem[\protect\citeauthoryear{Kenjere{\v{s}} and Hanjali{\'c}}{2000}]{Kenjere1997}
\begin{barticle}
\bauthor{\bsnm{Kenjere{\v{s}}}, \binits{S.}},
\bauthor{\bsnm{Hanjali{\'c}}, \binits{K.}}:
\batitle{On the implementation of effects of lorentz force in turbulence closure models}.
\bjtitle{International Journal of Heat and Fluid Flow}
\bvolume{21}(\bissue{3}),
\bfpage{329}--\blpage{337}
(\byear{2000})
\end{barticle}
\endbibitem

\bibitem[\protect\citeauthoryear{Lee and Choi}{2001}]{Lee2001}
\begin{barticle}
\bauthor{\bsnm{Lee}, \binits{D.}},
\bauthor{\bsnm{Choi}, \binits{H.}}:
\batitle{Magnethodrodynamic turbulent flow in a channel at low magnetic reynolds number}.
\bjtitle{Journal of Fluid Mechanics}
\bvolume{439},
\bfpage{367}--\blpage{394}
(\byear{2001})
\doiurl{10.1017/S0022112001004621}
\end{barticle}
\endbibitem

\bibitem[\protect\citeauthoryear{Ling et~al.}{2016}]{Ling2016}
\begin{barticle}
\bauthor{\bsnm{Ling}, \binits{J.}},
\bauthor{\bsnm{Kurzawski}, \binits{A.}},
\bauthor{\bsnm{Templeton}, \binits{J.}}:
\batitle{Reynolds averaged turbulence modelling using deep neural networks with embedded invariance}.
\bjtitle{Journal of Fluid Mechanics}
\bvolume{807},
\bfpage{155}--\blpage{166}
(\byear{2016})
\doiurl{10.1017/jfm.2016.615}
\end{barticle}
\endbibitem

\bibitem[\protect\citeauthoryear{Lundberg and Lee}{2017}]{NIPS2017_7062}
\begin{bchapter}
\bauthor{\bsnm{Lundberg}, \binits{S.M.}},
\bauthor{\bsnm{Lee}, \binits{S.-I.}}:
\bctitle{A unified approach to interpreting model predictions}.
In: \beditor{\bsnm{Guyon}, \binits{I.}},
\beditor{\bsnm{Luxburg}, \binits{U.V.}},
\beditor{\bsnm{Bengio}, \binits{S.}},
\beditor{\bsnm{Wallach}, \binits{H.}},
\beditor{\bsnm{Fergus}, \binits{R.}},
\beditor{\bsnm{Vishwanathan}, \binits{S.}},
\beditor{\bsnm{Garnett}, \binits{R.}} (eds.)
\bbtitle{Advances in Neural Information Processing Systems 30},
pp. \bfpage{4765}--\blpage{4774}.
\bpublisher{Curran Associates, Inc.},
\blocation{Long Beach, California, USA}
(\byear{2017}).
\burl{http://papers.nips.cc/paper/7062-a-unified-approach-to-interpreting-model-predictions.pdf}
\end{bchapter}
\endbibitem

\bibitem[\protect\citeauthoryear{Lumley}{1979}]{lumley1979computational}
\begin{barticle}
\bauthor{\bsnm{Lumley}, \binits{J.L.}}:
\batitle{Computational modeling of turbulent flows}.
\bjtitle{Advances in applied mechanics}
\bvolume{18},
\bfpage{123}--\blpage{176}
(\byear{1979})
\end{barticle}
\endbibitem

\bibitem[\protect\citeauthoryear{Mistrangelo et~al.}{2021}]{MISTRANGELO2021112795}
\begin{barticle}
\bauthor{\bsnm{Mistrangelo}, \binits{C.}},
\bauthor{\bsnm{Bühler}, \binits{L.}},
\bauthor{\bsnm{Smolentsev}, \binits{S.}},
\bauthor{\bsnm{Klüber}, \binits{V.}},
\bauthor{\bsnm{Maione}, \binits{I.}},
\bauthor{\bsnm{Aubert}, \binits{J.}}:
\batitle{Mhd flow in liquid metal blankets: Major design issues, mhd guidelines and numerical analysis}.
\bjtitle{Fusion Engineering and Design}
\bvolume{173},
\bfpage{112795}
(\byear{2021})
\doiurl{10.1016/j.fusengdes.2021.112795}
\end{barticle}
\endbibitem

\bibitem[\protect\citeauthoryear{Menter}{1994}]{menter1994two}
\begin{barticle}
\bauthor{\bsnm{Menter}, \binits{F.R.}}:
\batitle{Two-equation eddy-viscosity turbulence models for engineering applications}.
\bjtitle{AIAA journal}
\bvolume{32}(\bissue{8}),
\bfpage{1598}--\blpage{1605}
(\byear{1994})
\end{barticle}
\endbibitem

\bibitem[\protect\citeauthoryear{Mandler and Weigand}{2022}]{Mandler2022}
\begin{botherref}
\oauthor{\bsnm{Mandler}, \binits{H.}},
\oauthor{\bsnm{Weigand}, \binits{B.}}:
A realizable and scale-consistent data-driven non-linear eddy viscosity modeling framework for arbitrary regression algorithms.
International Journal of Heat and Fluid Flow
\textbf{97}
(2022)
\doiurl{10.1016/j.ijheatfluidflow.2022.109018}
\end{botherref}
\endbibitem

\bibitem[\protect\citeauthoryear{Miro et~al.}{2023}]{Miro2023}
\begin{bchapter}
\bauthor{\bsnm{Miro}, \binits{A.}},
\bauthor{\bsnm{Wallin}, \binits{S.}},
\bauthor{\bsnm{Colombo}, \binits{A.}},
\bauthor{\bsnm{Temmerman}, \binits{L.}},
\bauthor{\bsnm{Wunsch}, \binits{D.}},
\bauthor{\bsnm{Lehmkuhl}, \binits{O.}}:
\bctitle{Towards a machine learning model for explicit algebraic reynolds stress modelling using multi-expression programming}.
In: \bbtitle{Proceedings of the 14th International ERCOFTAC Symposium on Engineering, Turbulence, Modelling and Measurements: 6th-8th September 2023, Barcelona, Spain}.
\bpublisher{European Research Community on Flow, Turbulence, and Conbustion (ERCOFTAC)},
\blocation{Barcelona}
(\byear{2023}).
\burl{http://hdl.handle.net/2117/394814}
\end{bchapter}
\endbibitem

\bibitem[\protect\citeauthoryear{Nicoud and Ducros}{1999}]{nicoud1999subgrid}
\begin{barticle}
\bauthor{\bsnm{Nicoud}, \binits{F.}},
\bauthor{\bsnm{Ducros}, \binits{F.}}:
\batitle{Subgrid-scale stress modelling based on the square of the velocity gradient tensor}.
\bjtitle{Flow, turbulence and Combustion}
\bvolume{62}(\bissue{3}),
\bfpage{183}--\blpage{200}
(\byear{1999})
\end{barticle}
\endbibitem

\bibitem[\protect\citeauthoryear{Parish and Duraisamy}{2016}]{Parish2016}
\begin{barticle}
\bauthor{\bsnm{Parish}, \binits{E.J.}},
\bauthor{\bsnm{Duraisamy}, \binits{K.}}:
\batitle{A paradigm for data-driven predictive modeling using field inversion and machine learning}.
\bjtitle{Journal of Computational Physics}
\bvolume{305},
\bfpage{758}--\blpage{774}
(\byear{2016})
\doiurl{10.1016/j.jcp.2015.11.012}
\end{barticle}
\endbibitem

\bibitem[\protect\citeauthoryear{Paszke et~al.}{2019}]{paszke2019pytorch}
\begin{botherref}
\oauthor{\bsnm{Paszke}, \binits{A.}},
\oauthor{\bsnm{Gross}, \binits{S.}},
\oauthor{\bsnm{Massa}, \binits{F.}},
\oauthor{\bsnm{Lerer}, \binits{A.}},
\oauthor{\bsnm{Bradbury}, \binits{J.}},
\oauthor{\bsnm{Chanan}, \binits{G.}},
\oauthor{\bsnm{Killeen}, \binits{T.}},
\oauthor{\bsnm{Lin}, \binits{Z.}},
\oauthor{\bsnm{Gimelshein}, \binits{N.}},
\oauthor{\bsnm{Antiga}, \binits{L.}}, et al.:
Pytorch: An imperative style, high-performance deep learning library.
Advances in neural information processing systems
\textbf{32}
(2019)
\end{botherref}
\endbibitem

\bibitem[\protect\citeauthoryear{Pope}{1975}]{Pope1975}
\begin{barticle}
\bauthor{\bsnm{Pope}, \binits{S.B.}}:
\batitle{A more general effective-viscosity hypothesis}.
\bjtitle{Journal of Fluid Mechanics}
\bvolume{72},
\bfpage{331}--\blpage{340}
(\byear{1975})
\doiurl{10.1017/S0022112075003382}
\end{barticle}
\endbibitem

\bibitem[\protect\citeauthoryear{Parashar et~al.}{2020}]{parashar2020modeling}
\begin{barticle}
\bauthor{\bsnm{Parashar}, \binits{N.}},
\bauthor{\bsnm{Srinivasan}, \binits{B.}},
\bauthor{\bsnm{Sinha}, \binits{S.S.}}:
\batitle{Modeling the pressure-hessian tensor using deep neural networks}.
\bjtitle{Physical Review Fluids}
\bvolume{5}(\bissue{11}),
\bfpage{114604}
(\byear{2020})
\end{barticle}
\endbibitem

\bibitem[\protect\citeauthoryear{Rubel}{2019}]{articlerubentritium}
\begin{barticle}
\bauthor{\bsnm{Rubel}, \binits{M.}}:
\batitle{Fusion neutrons: Tritium breeding and impact on wall materials and components of diagnostic systems}.
\bjtitle{Journal of Fusion Energy}
\bvolume{38},
\bfpage{1}--\blpage{15}
(\byear{2019})
\doiurl{10.1007/s10894-018-0182-1}
\end{barticle}
\endbibitem

\bibitem[\protect\citeauthoryear{Schmelzer et~al.}{2020}]{Schmelzer2020}
\begin{barticle}
\bauthor{\bsnm{Schmelzer}, \binits{M.}},
\bauthor{\bsnm{Dwight}, \binits{R.P.}},
\bauthor{\bsnm{Cinnella}, \binits{P.}}:
\batitle{Discovery of algebraic reynolds-stress models using sparse symbolic regression}.
\bjtitle{Flow, Turbulence and Combustion}
\bvolume{104},
\bfpage{579}--\blpage{603}
(\byear{2020})
\doiurl{10.1007/s10494-019-00089-x}
\end{barticle}
\endbibitem

\bibitem[\protect\citeauthoryear{Saez-de Ocáriz-Borde et~al.}{2021}]{Borde2021}
\begin{botherref}
\oauthor{\bsnm{Saez-de-Ocáriz-Borde}, \binits{H.}},
\oauthor{\bsnm{Sondak}, \binits{D.}},
\oauthor{\bsnm{Protopapas}, \binits{P.}}:
Multi-task learning based convolutional models with curriculum learning for the anisotropic reynolds stress tensor in turbulent duct flow.
ArXiv, abs/2111.00328
(2021)
\end{botherref}
\endbibitem

\bibitem[\protect\citeauthoryear{Steiner et~al.}{2020}]{Steiner2020}
\begin{bchapter}
\bauthor{\bsnm{Steiner}, \binits{J.}},
\bauthor{\bsnm{Dwight}, \binits{R.}},
\bauthor{\bsnm{Vir{\'e}}, \binits{A.}}:
\bctitle{Data-driven turbulence modeling for wind turbine wakes under neutral conditions}.
In: \bbtitle{Journal of Physics: Conference Series},
vol. \bseriesno{1618},
p. \bfpage{062051}
(\byear{2020}).
\bcomment{IOP Publishing}
\end{bchapter}
\endbibitem

\bibitem[\protect\citeauthoryear{Smolentsev}{2021}]{smolentsev2021physical}
\begin{barticle}
\bauthor{\bsnm{Smolentsev}, \binits{S.}}:
\batitle{Physical background, computations and practical issues of the magnetohydrodynamic pressure drop in a fusion liquid metal blanket}.
\bjtitle{Fluids}
\bvolume{6}(\bissue{3}),
\bfpage{110}
(\byear{2021})
\end{barticle}
\endbibitem

\bibitem[\protect\citeauthoryear{Weller et~al.}{1998}]{openfoam}
\begin{barticle}
\bauthor{\bsnm{Weller}, \binits{H.G.}},
\bauthor{\bsnm{Tabor}, \binits{G.}},
\bauthor{\bsnm{Jasak}, \binits{H.}},
\bauthor{\bsnm{Fureby}, \binits{C.}}:
\batitle{A tensorial approach to computational continuum mechanics using object-oriented techniques}.
\bjtitle{Computers in physics}
\bvolume{12}(\bissue{6}),
\bfpage{620}--\blpage{631}
(\byear{1998})
\end{barticle}
\endbibitem

\bibitem[\protect\citeauthoryear{Wang et~al.}{2017}]{Wang2017}
\begin{botherref}
\oauthor{\bsnm{Wang}, \binits{J.X.}},
\oauthor{\bsnm{Wu}, \binits{J.L.}},
\oauthor{\bsnm{Xiao}, \binits{H.}}:
Physics-informed machine learning approach for reconstructing reynolds stress modeling discrepancies based on dns data.
Physical Review Fluids
\textbf{2}
(2017)
\doiurl{10.1103/PhysRevFluids.2.034603}
\end{botherref}
\endbibitem

\bibitem[\protect\citeauthoryear{Wu et~al.}{2018}]{Wu2018}
\begin{botherref}
\oauthor{\bsnm{Wu}, \binits{J.L.}},
\oauthor{\bsnm{Xiao}, \binits{H.}},
\oauthor{\bsnm{Paterson}, \binits{E.}}:
Physics-informed machine learning approach for augmenting turbulence models: A comprehensive framework.
Physical Review Fluids
\textbf{7}
(2018)
\doiurl{10.1103/PhysRevFluids.3.074602}
\end{botherref}
\endbibitem

\bibitem[\protect\citeauthoryear{Zhang et~al.}{2019}]{Zhang2019}
\begin{barticle}
\bauthor{\bsnm{Zhang}, \binits{Z.}},
\bauthor{\bsnm{Song}, \binits{X.}},
\bauthor{\bsnm{Ye}, \binits{S.}},
\bauthor{\bsnm{Wang}, \binits{Y.}},
\bauthor{\bsnm{Huang}, \binits{C.}},
\bauthor{\bsnm{An}, \binits{Y.}},
\bauthor{\bsnm{Chen}, \binits{Y.}}:
\batitle{Application of deep learning method to reynolds stress models of channel flow based on reduced-order modeling of dns data}.
\bjtitle{Journal of Hydrodynamics}
\bvolume{31},
\bfpage{58}--\blpage{65}
(\byear{2019})
\doiurl{10.1007/s42241-018-0156-9}
\end{barticle}
\endbibitem

\end{thebibliography}

\end{document}